\documentclass{sig-alternate-2013}

\usepackage[
  pass,
]{geometry}

\usepackage{rotating}
\usepackage{subfigure}
\usepackage{multirow}
\usepackage{color}
\usepackage{verbatim}

\usepackage{tikz}
\usetikzlibrary{shapes,shadows,arrows,backgrounds}
\usetikzlibrary[arrows,decorations.pathmorphing,backgrounds,positioning,fit,petri]
\usepgfmodule{plot}

\begin{document}

\newfont{\mycrnotice}{ptmr8t at 7pt}
\newfont{\myconfname}{ptmri8t at 7pt}
\let\crnotice\mycrnotice%
\let\confname\myconfname%

\permission{Publication rights licensed to ACM. ACM acknowledges that this contribution was authored or co-authored by an employee, contractor or affiliate of a national government. As such, the Government retains a nonexclusive, royalty-free right to publish or reproduce this article, or to allow others to do so, for Government purposes only.}
\conferenceinfo{DOLAP'14,}{November 7, 2014, Shanghai, China.}
\copyrightetc{Copyright is held by the owner/author(s). Publication rights licensed to ACM.\\
ACM 978-1-4503-0999-8/14/11\ ...\$15.00.\\
http://dx.doi.org/10.1145/2666158.2666173}

\clubpenalty=10000
\widowpenalty = 10000

\title{fVSS: A New Secure and Cost-Efficient Scheme \\for Cloud Data Warehouses}

\numberofauthors{3} 
\author{
\alignauthor
Varunya Attasena\\
       \affaddr{Université de Lyon (ERIC)}\\
       \affaddr{Université Lumière Lyon 2}\\
       \affaddr{5 av. Pierre Mendès-France}\\
       \affaddr{69676 Bron Cedex -- France}\\
       \email{vattasena@eric.univ-lyon2.fr}
\alignauthor
Nouria Harbi\\
       \affaddr{Université de Lyon (ERIC)}\\
       \affaddr{Université Lumière Lyon 2}\\
       \affaddr{5 av. Pierre Mendès-France}\\
       \affaddr{69676 Bron Cedex -- France}\\
       \email{nouria.harbi@univ-lyon2.fr}
\alignauthor 
Jérôme Darmont\\
       \affaddr{Université de Lyon (ERIC)}\\
       \affaddr{Université Lumière Lyon 2}\\
       \affaddr{5 av. Pierre Mendès-France}\\
       \affaddr{69676 Bron Cedex -- France}\\
       \email{jerome.darmont@univ-lyon2.fr}}

\maketitle

\begin{abstract}
Cloud business intelligence is an increasingly popular choice to deliver decision support capabilities via elastic, pay-per-use resources. However, data security issues are one of the top concerns when dealing with sensitive data.
In this paper, we propose a novel approach for securing cloud data warehouses by flexible verifiable secret sharing, fVSS. Secret sharing encrypts and distributes data over several cloud service providers, thus enforcing data privacy and availability. fVSS addresses four shortcomings in existing secret sharing-based approaches. First, it allows refreshing the data warehouse when some service providers fail. Second, it allows on-line analysis processing. Third, it enforces data integrity with the help of both inner and outer signatures. Fourth, it helps users control the cost of cloud warehousing by balancing the load among service providers with respect to their pricing policies.
To illustrate fVSS' efficiency, we thoroughly compare it with existing secret sharing-based approaches with respect to security features, querying power and data storage and computing costs. 
\end{abstract}

\category{H. Information Systems}{H.2. Data Management}{H.2.7. Database administration}[Data Warehouse and Repository; Security, Integrity and Protection]



\section{Introduction}

Cloud business intelligence (BI) is becoming increasingly popular, providing the benefits of both classical BI (efficient decision-support) and cloud computing (elasticity of resources and costs). However, one of the top concerns of cloud users and would-be users remains security. Some security issues are inherited from classical distributed architectures, e.g., authentication, network attacks and vulnerability exploitation, but some directly relate to the new framework of the cloud, e.g., cloud service provider (CSP) or subcontractor espionage, cost-effective defense of availability and uncontrolled mashups \cite{Chow09}.
In this paper, we focus on data security, which is of critical importance in a BI context where sensitive data are processed. Data security is usually managed by CSPs, but with the multiplication of CSPs and subcontractors in many different countries, intricate legal issues arise and trust in CSPs may be jeopardized. 


In contrast, end-to-end security gives the control of both data security levels and costs back to users, through classical techniques such as data encryption, anonymization, replication and verification. However, updating and querying secured data in the cloud may turn inefficient and expensive. Thus, specific end-to-end security approaches were designed for distributed/cloud databases (DBs) and data warehouses (DWs) \cite{Ravan13,Sion08}. The MONOMI system \cite{Tu13} notably encrypts both SQL queries and data with multiple cryptographic schemes. However, although MONOMI enforces data privacy, it addresses neither data availability nor integrity issues. In contrast, secret sharing \cite{SSS:Shamir-1979} simultaneously enforces data privacy, availability and integrity. Secret sharing transforms sensitive data into individually meaningless data pieces (called shares) that are distributed to $n$ CSPs. Computations can then be performed onto shares, but yield meaningless individual results. The global result can only be reconstructed (decrypted) by the user.



All secret sharing-based approaches allow accessing shares from $t \leq n$ CSPs, i.e., shares are still available when up to $n - t$ CSPs fail, e.g., go bankruptcy, due to a technical problem or even by malice. However, no approach allows updating shares when even one single CSP fails, thus hindering cloud DWs' refreshment capabilities. Moreover, although these approaches feature all basic DB querying operators, only one handles on-line analysis processing (OLAP) and it does not support lazy updates on cloud-stored cubes. In addition, few approaches actually enforce data integrity through verification of inner code (to verify whether CSPs are malicious) or outer code (to detect incorrect data before decryption), and only one features both inner and outer signatures for this sake. Finally, although some approaches bring in solutions to reduce overall storage volume so that it falls well under $n$ times that of original data, and thus decrease monetary cost in the pay-as-you-go paradigm, there is still room for improvement.


To address all these issues, we propose a novel approach that relies on a flexible verifiable secret sharing scheme (fVSS). To the best of our knowledge, fVSS is the first approach allowing DW refreshment when one or several CSPs fail. Moreover, fVSS allows running OLAP operators on shared DWs or cubes without reconstructing all data first. fVSS also features both inner and outer signatures for data verification. Finally, fVSS allows users adjusting the volume of shared data at each CSP, which helps optimize cost with respect to various CFP pricing policies. 

The remainder of this paper is organized as follows. Section~\ref{sec:Related works} discusses previous research related to fVSS. Section~\ref{sec:our-approach} details our secret sharing and reconstruction mechanisms, the new outer signature we propose and how our approach applies to data warehouses and OLAP. Section~\ref{sec:Comparison-approaches} provides a comparative study of fVSS with state-of-the-art existing approaches. Finally, Section~\ref{sec:conclusion} concludes this paper and hints at future research perspectives.

\section{Related works} 
\label{sec:Related works}


Among encryption techniques, only secret sharing \cite{Survey-SS} handles both data privacy and availability, which is why we focus on this family of approaches. The principle of secret sharing \cite{SSS:Shamir-1979} is based on the fact that $t$ points define a polynomial $y=f(x)$ of degree $t-1$. The secret is the polynomial's constant term and the remaining terms are usually randomly selected. Each data piece is transformed into $n$ shares $f(x_i)$ corresponding to points of the polynomial.
Reconstruction of the secret is achieved through Lagrange interpolation \cite{Lagrange}: there is only one polynomial $p(x)$ such that $degree(p(x)) < t$ and $p(x_i) = f(x_i)$. Then the secret is $p(0)$.
Moreover, modern secret sharing schemes, such as multi-secret sharing \cite{DB:Attasena-et-al-2013,MSSS:Liu12,MSSS:Waseda12}, verifiable secret sharing \cite{VSSS:Bu09}, and verifiable multi-secret sharing \cite{VMSSS:Bu12,VMSSS:Eslami10}, also help reduce shared data volume, verify the honesty of CSPs, and both, respectively. 
We classify secret sharing-based approaches for securing DBs and DWs into two families.

\begin{figure}[hbt]
	
	\resizebox{85mm}{!} {
		\begin{tikzpicture}
		
			\node [rectangle,draw,dashed,minimum size=3.3cm,minimum width=4.8cm] at (1.95,0) (data) {$\ $};	
			\node at (0.1,1.5) {\textsf{\tiny{Data owner}}};
		
			\node [rectangle,fill=black!100,minimum size=5mm,rounded corners=2mm,minimum width=4mm] at (1.3,0) (process) {\color{white} \textsf{\textbf{\small{$\left( t,n\right) $}\scriptsize{ SSS}}}};
			
			\node [rectangle,draw,thick,minimum size=7mm,minimum width=5mm] at (0,0) (data) {\textbf{\small{$T_{j}$}}};
			\node at (0,-0.5) {\textsf{\tiny{Original}}};
			\node at (0,-0.7) {\textsf{\tiny{table}}};
			\draw [-stealth,thick] (data) -- (process);
			\draw [-stealth,thick,gray] (0.55,0.1) -- (0.25,0.1);
						
			\node [rectangle,draw,thick,minimum size=7mm,minimum width=6.5mm] at (3,1.2) (E1) {~};
			\node at (3,1.2) {\textbf{\tiny{$ET_{j1}$}}};
			\node [rectangle,draw,thick,minimum size=7mm,minimum width=6.5mm] at (3,0) (Et) {~};
			\node at (3,0) {\textbf{\tiny{$ET_{jt}$}}};	
			\node [rectangle,draw,thick,minimum size=7mm,minimum width=6.5mm] at (3,-1.2) (En) {~};
			\node at (3,-1.2) {\textbf{\tiny{$ET_{jn}$}}};
			
			\node at (3, 0.7) {\textsf{\tiny{$\vdots$}}};
			\node at (3,-0.5) {\textsf{\tiny{$\vdots$}}};
			
			\node at (3.68,1.3) {\textsf{\tiny{Shared}}};
			\node at (3.7,1.1) {\textsf{\tiny{table 1}}};
			\node at (3.68,0.1) {\textsf{\tiny{Shared}}};
			\node at (3.7,-0.1) {\textsf{\tiny{table $t$}}};
			\node at (3.68,-1.1) {\textsf{\tiny{Shared}}};
			\node at (3.7,-1.3) {\textsf{\tiny{table $n$}}};
			
			\draw [-stealth,thick] (2.05,0) -- (2.65,1.2);
			\draw [-stealth,thick] (process) -- (Et);
			\draw [-stealth,thick] (2.05,0) -- (2.65,-1.2);
			\draw [-stealth,thick, gray] (2.65,1.4) -- (2.05,0.2);
			\draw [-stealth,thick,gray] (2.65,0.1) -- (2.1,0.1);

			\node [rectangle,draw,thick,minimum size=7mm,minimum width=6.5mm] at (5.3,1.2) (E1-CSP) {~};
			\node at (5.3, 1.2) {\textbf{\tiny{$ET_{j1}$}}};	
			\node [rectangle,draw,thick,minimum size=7mm,minimum width=6.5mm] at (5.3,0) (Et-CSP) {~};
			\node at (5.3, 0.0) {\textbf{\tiny{$ET_{jt}$}}};	
			\node [rectangle,draw,thick,minimum size=7mm,minimum width=6.5mm] at (5.3,-1.2) (En-CSP) {~};
			\node at (5.3,-1.2) {\textbf{\tiny{$ET_{jn}$}}};
			
			\node at (5.3, 0.7) {\textsf{\tiny{$\vdots$}}};
			\node at (5.3,-0.5) {\textsf{\tiny{$\vdots$}}};
			
			\draw [-stealth,thick] (4.2,1.2) -- (E1-CSP);
			\draw [-stealth,thick] (4.2,0) -- (Et-CSP);
			\draw [-stealth,thick] (4.2,-1.2) -- (En-CSP);
			\draw [-stealth,thick,gray] (4.9,1.3) -- (4.2,1.3);
			\draw [-stealth,thick,gray] (4.9,0.1) -- (4.2,0.1);
			
			\node at (6.32,1.3) {\textsf{\tiny{Shared table 1}}};
			\node at (6.2,1.1) {\textsf{\tiny{at $CSP_1$}}};
			\node at (6.32,0.1) {\textsf{\tiny{Shared table $t$}}};
			\node at (6.2,-0.1) {\textsf{\tiny{at $CSP_t$}}};
			\node at (6.32,-1.1) {\textsf{\tiny{Shared table $n$}}};
			\node at (6.2,-1.3) {\textsf{\tiny{at $CSP_n$}}};

			\draw [-stealth,thick] (-0.3,-1.3) -- (0,-1.3);
			\draw [-stealth,thick,gray] (0,-1.5) -- (-0.3,-1.5);
			\node at (0.6,-1.3) {\textsf{\tiny{Sharing process}}};
			\node at (0.9,-1.5) {\textsf{\tiny{Reconstruction process}}};
				
		\end{tikzpicture}
	} 
	\vspace{-0.5cm}	
	\caption{First strategy for sharing a database}
	\label{fig:1st-DB-framework}
\end{figure}
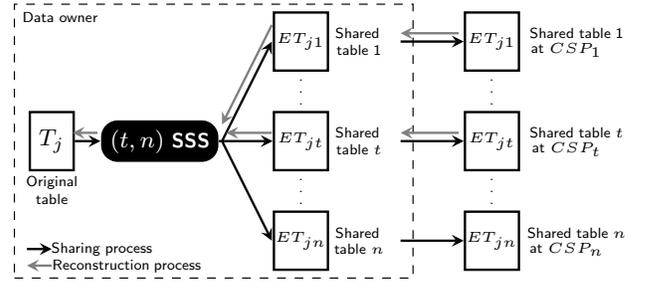

In the first family of approaches \cite{DB:Agrawal-et-al-2009,DB:Attasena-et-al-2013,DB:Emekci-et-al-2005,DB:Emekci-et-al-2006,DB:Hadavi-et-al-2013}, each table is encrypted into $n$ shared tables, each of which is stored at one given CSP (Figure \ref{fig:1st-DB-framework}). Recall that only $t$ of $n$ shared tables are sufficient to reconstruct the original table. Most of these approaches assume that CSPs are not malicious and that connections between CSPs and users are secure. Only one \cite{DB:Attasena-et-al-2013} includes a data verification process that exploits hash-generated signatures: an inner signature (incorporated to the shares) to verify whether CSPs are malicious, and an outer signature that helps detect incorrect or erroneous (lost, damaged, alternative...) data before decryption and prevents useless data transfers. Both signatures are stored at CSPs.


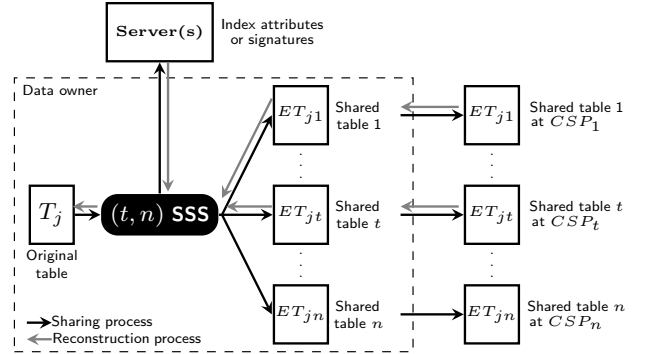
\begin{figure}[hbt]
	\resizebox{85mm}{!} {
		\begin{tikzpicture}
		
			\node [rectangle,draw,dashed,minimum size=3.3cm,minimum width=4.8cm] at (1.95,0) (data) {$\ $};	
			\node at (0.1,1.5) {\textsf{\tiny{Data owner}}};
		
			\node [rectangle,fill=black!100,minimum size=5mm,rounded corners=2mm,minimum width=4mm] at (1.3,0) (process) {\color{white} \textsf{\textbf{\small{$\left( t,n\right) $}\scriptsize{ SSS}}}};
			
			\node [rectangle,draw,thick,minimum size=7mm,minimum width=5mm] at (0,0) (data) {\textbf{\small{$T_{j}$}}};
			\node at (0,-0.5) {\textsf{\tiny{Original}}};
			\node at (0,-0.7) {\textsf{\tiny{table}}};
			\draw [-stealth,thick] (data) -- (process);
			\draw [-stealth,thick,gray] (0.55,0.1) -- (0.25,0.1);
						
			\node [rectangle,draw,thick,minimum size=7mm,minimum width=6.5mm] at (3,1.2) (E1) {~};
			\node at (3,1.2) {\textbf{\tiny{$ET_{j1}$}}};
			\node [rectangle,draw,thick,minimum size=7mm,minimum width=6.5mm] at (3,0) (Et) {~};
			\node at (3,0) {\textbf{\tiny{$ET_{jt}$}}};	
			\node [rectangle,draw,thick,minimum size=7mm,minimum width=6.5mm] at (3,-1.2) (En) {~};
			\node at (3,-1.2) {\textbf{\tiny{$ET_{jn}$}}};
			
			\node at (3, 0.7) {\textsf{\tiny{$\vdots$}}};
			\node at (3,-0.5) {\textsf{\tiny{$\vdots$}}};
			
			\node at (3.68,1.3) {\textsf{\tiny{Shared}}};
			\node at (3.7,1.1) {\textsf{\tiny{table 1}}};
			\node at (3.68,0.1) {\textsf{\tiny{Shared}}};
			\node at (3.7,-0.1) {\textsf{\tiny{table $t$}}};
			\node at (3.68,-1.1) {\textsf{\tiny{Shared}}};
			\node at (3.7,-1.3) {\textsf{\tiny{table $n$}}};
			
			\draw [-stealth,thick] (2.05,0) -- (2.65,1.2);
			\draw [-stealth,thick] (process) -- (Et);
			\draw [-stealth,thick] (2.05,0) -- (2.65,-1.2);
			\draw [-stealth,thick, gray] (2.65,1.4) -- (2.05,0.2);
			\draw [-stealth,thick,gray] (2.65,0.1) -- (2.1,0.1);

			\node [rectangle,draw,thick,minimum size=7mm,minimum width=6.5mm] at (5.3,1.2) (E1-CSP) {~};
			\node at (5.3, 1.2) {\textbf{\tiny{$ET_{j1}$}}};	
			\node [rectangle,draw,thick,minimum size=7mm,minimum width=6.5mm] at (5.3,0) (Et-CSP) {~};
			\node at (5.3, 0.0) {\textbf{\tiny{$ET_{jt}$}}};	
			\node [rectangle,draw,thick,minimum size=7mm,minimum width=6.5mm] at (5.3,-1.2) (En-CSP) {~};
			\node at (5.3,-1.2) {\textbf{\tiny{$ET_{jn}$}}};
			
			\node at (5.3, 0.7) {\textsf{\tiny{$\vdots$}}};
			\node at (5.3,-0.5) {\textsf{\tiny{$\vdots$}}};
			
			\draw [-stealth,thick] (4.2,1.2) -- (E1-CSP);
			\draw [-stealth,thick] (4.2,0) -- (Et-CSP);
			\draw [-stealth,thick] (4.2,-1.2) -- (En-CSP);
			\draw [-stealth,thick,gray] (4.9,1.3) -- (4.2,1.3);
			\draw [-stealth,thick,gray] (4.9,0.1) -- (4.2,0.1);
			
			\node at (6.32,1.3) {\textsf{\tiny{Shared table 1}}};
			\node at (6.2,1.1) {\textsf{\tiny{at $CSP_1$}}};
			\node at (6.32,0.1) {\textsf{\tiny{Shared table $t$}}};
			\node at (6.2,-0.1) {\textsf{\tiny{at $CSP_t$}}};
			\node at (6.32,-1.1) {\textsf{\tiny{Shared table $n$}}};
			\node at (6.2,-1.3) {\textsf{\tiny{at $CSP_n$}}};

			\node [rectangle,draw,thick,minimum size=7mm,minimum width=5mm] at (1.3,2.2) (F-index) {\textbf{\tiny{Server(s)}}};
			\node at (2.65,2.3) {\textsf{\tiny{Index attributes}}};
			\node at (2.65,2.1) {\textsf{\tiny{or signatures}}};
			\draw [-stealth,thick] (process) -- (F-index);
			\draw [-stealth,thick,gray] (1.4,1.83) -- (1.4,0.28);

			\draw [-stealth,thick] (-0.3,-1.3) -- (0,-1.3);
			\draw [-stealth,thick,gray] (0,-1.5) -- (-0.3,-1.5);
			\node at (0.6,-1.3) {\textsf{\tiny{Sharing process}}};
			\node at (0.9,-1.5) {\textsf{\tiny{Reconstruction process}}};
				
		\end{tikzpicture}
	} 
	\vspace{-0.5cm}	
	\caption{Second strategy for sharing a database}	
	\label{fig:2nd-DB-framework}
\end{figure}

In the second family of approaches \cite{DB:Hadavi-Jalili-2010,DB:Hadavi-et-al-2012,DB:Thompson-et-al-2009,DB:Wang-et-al-2011}, one or more additional index servers, located at $\{CSP_i\}_{i>n}$,  store B++ tree indices and signatures (Figure~\ref{fig:2nd-DB-framework}). The index servers require higher security and computing power than that of other nodes, and a secure connection to the user's. The index servers support data verification by various means, i.e., homomorphic encryption \cite{DB:Hadavi-et-al-2012}, a hash function \cite{DB:Thompson-et-al-2009} and  checksums and a hash function \cite{DB:Wang-et-al-2011}. However, data verification cannot take place if the index server fails.


Regarding accessing shares, classical secret sharing \cite{SSS:Shamir-1979} and most of its above-cited extensions natively support some exact match and aggregation operators, i.e., equality and inequality, sum, average and count. Each approach handles operators needing sorted data, e.g., range operators, maximum and minimum, with various techniques: preaggregation before sharing \cite{DB:Emekci-et-al-2005}, a B++ tree index \cite{DB:Hadavi-Jalili-2010,DB:Hadavi-et-al-2012} or the rank coefficient used in classical secret sharing \cite{DB:Agrawal-et-al-2009,DB:Emekci-et-al-2006,DB:Hadavi-et-al-2013}. However, extra storage space is needed to store these data structures. Finally, almost all approaches allow updates on shares, since each piece of data is encrypted independently.

The features of all above-cited approaches are summarized in Table~\ref{tab:Comparison-approaches}. We compare them with fVSS' in Section~\ref{sec:Comparison-approaches}.

\section{Flexible Verifiable Secret \\Sharing} 
\label{sec:our-approach}

\subsection{fVSS Principle} 

fVSS is a $(t, n)$ flexible verifiable secret sharing scheme belonging to the second family of approaches identified in Section~\ref{sec:Related works}. As all similar approaches, fVSS shares data over $n$ CSPs, $t$ of which are necessary to reconstruct original data. Table~\ref{tab:Parameters-setting} lists fVSS' parameters, which will be introduced throughout this section.

\begin{table}[hbt]
\centering
    \caption{fVSS parameters}     
    \label{tab:Parameters-setting}
   	\resizebox{85mm}{!} {
    \begin{tabular}{|c|l|}
    \hline
    {\bfseries Parameters} & \multicolumn{1} {c|} {\bfseries Definitions} \\ \hline
    $p$ & A big prime number\\ \hline
    $n$ & Number of CSPs \\ \hline
    $t$ & Number of shares necessary for reconstructing original data \\ \hline
    $CSP_i$ & CSP number $i$ \\ \hline
    $ID_i$ & Identifier number of $CSP_i$ such that $p>ID_i>1$ \\ \hline
    $m$ & Number of tables \\ \hline
    $T_j$ & Table number $j$ such that $T_j=\left\lbrace R_{j,k} \right\rbrace _{k=1...r_j} $\\ \hline
    $ET_{ij}$ & Shared table of $T_j$ stored at $CSP_i$\\ \hline
    $q_j$ & Number of attributes of $T_j$ and $ET_{ij}$ \\ & (not including primary key) \\ \hline
    $r_j$ & Number of records of $T_j$ \\ \hline
    $er_{ij}$ & Number of records of $ET_{ij}$ such that $r_j\geq er_{ij}$ \\ & and $\sum _{i=1}^n er_{ij}=\left(n-t+2 \right) r_j$ \\ \hline
    $A_{jl}$ & Attribute number $l$ of $T_j$ and $ET_{ij}$ \\ \hline
    $R_{jk}$ & Record $\#k$ of $T_j$ such that $R_{jk}=\left\lbrace pk_{jk},d_{jk1}\dots d_{jkq_j} \right\rbrace$ \\ \hline
    $ER_{ijg}$ & Record $\#g$ of $ET_{ij}$ such that $ER_{ijg}=\left\lbrace pk_{ijg},e_{ijg1}\dots e_{ijgq_j} \right\rbrace$ \\ 
    & $ER_{ijg}$ is shared record of $R_{jk}$ if $pk_{jk}=pk_{ijg}$ \\ \hline
    $pk_{jk}$ & Primary key value of $R_{jk}$ \\ \hline
    $pk_{ijg}$ & Primary key value of $ER_{ijg}$. It is not encrypted. \\ \hline
    $d_{jkl}$ & Value of $A_{jl}$ of $R_{jk}$ such that $p>d_{jkl}\geq0$ \\ \hline
    $e_{ijgl}$ & Share of $d_{jkl}$ stored in $A_{jl}$ of $ER_{ijg}$ such that $p>e_{ijgl}\geq0$. \\ \hline
    $SG_{jk}$ & Group of CSPs that store shares of $R_{jk}$ such that \\ & $ SG_{jk} \subset \lbrace CSP_i \rbrace_{i=1..n}$ and $\vert SG_{jk}\vert = n-t+2$ \\ \hline
    $UG_{jk}$ & Group of CSPs that do \emph{not} store shares of $R_{jk}$ such that\\ &  $UG_{jk}  \subset \lbrace CSP_i \rbrace_{i=1..n}$ and $UG_{jk} = \lbrace CSP_i \rbrace_{i=1..n}- SG_{jk}$ \\ \hline
    $RG$ & Group of CSPs selected to reconstruct data such that \\ &  $RG \subset \lbrace CSP_i \rbrace_{i=1..n}$ and $\vert RG\vert = t$ \\ \hline
    $s\_in_{jkl}$ & Inner signature of $d_{jkl}$ such that $p>s\_in_{jkl}\geq 0$ \\ \hline
    $s\_rout_{ijuv}$ & Outer record signature number $v$, level $u$ of $ER_{ijg}$ in $ET_{ij}$ \\ & stored at $CSP_i$ \\ \hline
    $s\_tout_{iuv}$ & Outer table signature number $v$, level $u$ of $ET_{ij}$ stored \\ & at $CSP_i$ \\ \hline
    $w_i$ & Maximum number of child nodes in $CSP_i$'s outer \\ &signature tree \\ \hline
    \end{tabular}
    }
\end{table}

The main novelty in fVSS is that, to optimize shared data volume and thus cost, we share a piece of data fewer than $n$ times. For example, in Figure \ref{fig:example-DW} where $n=5$, record \#124 of table PRODUCT is only shared at $CSP_1$, $CSP_3$ and $CSP_5$, which presumably feature the lowest storage costs. To achieve this, we proceed as follows. 


\begin{figure}[hbt]
\begin{center}
\subfigure[Original data]{ 
   \resizebox{4cm}{!} {
    \begin{tabular}{|c|c|c|c|c|}
    \hline
    {\bfseries ProNo} & {\bfseries ProName} & {\bfseries ProDescr} & {\bfseries CategoryID} & {\bfseries UnitPrice} \\ \hline    
    124 & Shirt & Red & 1 & 75 \\ \hline
    125 & Shoe & \textit{NULL} & 2 & 80 \\ \hline
    126 & Ring & \textit{NULL} & 1 & 80 \\ \hline
    \end{tabular}
    } 
 }\subfigure[Indices on index server]{
   \resizebox{4cm}{!} {
    \begin{tabular}{c|c|c|c}
    \cline{2-3}
    {$\hspace{75pt}$} & {\bfseries ProNo} & {\bfseries Share location} & {$\hspace{75pt}$}\\ \cline{2-3}    
    & 124 & 10101 & \\ \cline{2-3}
    & 125 & 01110 & \\ \cline{2-3}
    & 126 & 11010 & \\ \cline{2-3}
    \end{tabular}
    }
 }
 \subfigure[Shares at $CSP_{1}$]{
   \resizebox{4cm}{!} {
    \begin{tabular}{|c|c|c|c|c|}
    \hline
    {\bfseries ProNo} & {\bfseries ProName} & {\bfseries ProDescr} & {\bfseries CategoryID} & {\bfseries UnitPrice} \\ \hline
    124 & \{6,5,3,11,7\} & \{10,5,8\} & 1 & 6 \\ \hline
    126 & \{10,3,6,12\} & \textit{NULL} & 2 & 45 \\ \hline
    \end{tabular}
    }
 }\subfigure[Shares at $CSP_{2}$]{
   \resizebox{4cm}{!} {
    \begin{tabular}{|c|c|c|c|c|}
    \hline
    {\bfseries ProNo} & {\bfseries ProName} & {\bfseries ProDescr} & {\bfseries CategoryID} & {\bfseries UnitPrice} \\ \hline
    125 & \{6,5,4,5\} & \textit{NULL} & 2 & 5 \\ \hline
    126 & \{2,6,11,10\} & \textit{NULL} & 6 & 8 \\ \hline
    \end{tabular}
    }
 }
 \subfigure[Shares at $CSP_{3}$]{
   \resizebox{4cm}{!} {
    \begin{tabular}{|c|c|c|c|c|}
    \hline
    {\bfseries ProNo} & {\bfseries ProName} & {\bfseries ProDescr} & {\bfseries CategoryID} & {\bfseries UnitPrice} \\ \hline
    124 & \{6,6,5,7,9\} & \{12,8,1\} & 4 & 7 \\ \hline
    125 & \{6,5,8,3\} & \textit{NULL} & 9 & 11 \\ \hline
    \end{tabular}
    }
 }\subfigure[Shares at $CSP_{4}$]{
   \resizebox{4cm}{!} {
    \begin{tabular}{|c|c|c|c|c|}
    \hline
    {\bfseries ProNo} & {\bfseries ProName} & {\bfseries ProDescr} & {\bfseries CategoryID} & {\bfseries UnitPrice} \\ \hline
    125 & \{9,15,13,8\} & \textit{NULL} & 12 & 7 \\ \hline
    126 & \{2,7,6,9\} & \textit{NULL} & 12 & 1 \\ \hline
    \end{tabular}
    }
 }
 \subfigure[Shares at $CSP_{5}$]{
   \resizebox{4cm}{!} {
    \begin{tabular}{|c|c|c|c|c|}
    \hline
    {\bfseries ProNo} & {\bfseries ProName} & {\bfseries ProDescr} & {\bfseries CategoryID} & {\bfseries UnitPrice} \\ \hline
    124 & \{5,9,11,1,5\} & \{10,6,7\} & 8 & 13 \\ \hline
    \end{tabular}
    }
 }
\end{center}
\vspace{-0.5cm}
\caption{Sample original and shared data} \label{fig:example-DW}
\end{figure}

\begin{figure}[htb]
	\resizebox{85mm}{!} {
		\begin{tikzpicture}
	
			\draw[-stealth,thick] (-0.2,0) -- (8.4,0) node[right] {\textsf{\textbf{\small{$x$}}}};
			\draw[-stealth,thick] (0,-0.2) -- (0,3.8) node[above] {\textsf{\textbf{\small{$y=f_{jkl}(x)$}}}};
			
			\draw[color=black,ultra thick] plot [smooth] coordinates {(0,96/40) (1,121/40) (2,75/40) (3,17/40) (4,6/40) (5,101/40) (6,99/40) (7,59/40) (8,40/40)};
					
			\node at (1,-0.2) {\textsf{\tiny{1}}};
			\node at (2,-0.2) {\textsf{\tiny{2}}};
			\node at (3,-0.2) {\textsf{\tiny{3}}};
			\node at (4,-0.2) {\textsf{\tiny{4}}};
			\node at (5,-0.2) {\textsf{\tiny{5}}};
			\node at (6,-0.2) {\textsf{\tiny{6}}};
			\node at (7,-0.2) {\textsf{\tiny{7}}};
			\node at (8,-0.2) {\textsf{\tiny{8}}};
			\draw [thick] (1,0) -- (1,-0.05);
			\draw [thick] (2,0) -- (2,-0.05);
			\draw [thick] (3,0) -- (3,-0.05);
			\draw [thick] (4,0) -- (4,-0.05);
			\draw [thick] (5,0) -- (5,-0.05);
			\draw [thick] (6,0) -- (6,-0.05);
			\draw [thick] (7,0) -- (7,-0.05);
			\draw [thick] (8,0) -- (8,-0.05);
			
			\node [rotate=25] at (0.6,-0.6) {\color{red} \textsf{\textbf{\tiny{$HF_1(ID_2)$}}}};			
			\draw [thick,red,dashed] (1,0) -- (1,121/40);
			\node at (1,121/40) {\color{red} \textbf{\Huge{.}}};
			\node at (1,130/40) {\color{red} \textsf{\textbf{\small{$e_{2jgl}$}}}};	
			\node at (1,140/40) {\color{red} \textsf{\textbf{\scriptsize{Share}}}};	
			
			\node [rotate=25] at (1.6,-0.6) {\color{blue} \textsf{\textbf{\tiny{$HF_1(K_d)$}}}};		
			\draw [thick,blue,dashed] (2,0) -- (2,75/40);
			\node at (2,75/40) {\color{blue} \textbf{\Huge{.}}};
			\node at (2.2,85/40) {\color{blue} \textsf{\textbf{\small{$d_{jkl}$}}}};	
			\node at (2.2,95/40) {\color{blue} \textsf{\textbf{\scriptsize{Data}}}};	
			
			\node [rotate=25] at (2.6,-0.6) {\color{red} \textsf{\textbf{\tiny{$HF_1(ID_1)$}}}};		
			\draw [thick,red,dashed] (3,0) -- (3,17/40);
			\node at (3,17/40) {\color{red} \textbf{\Huge{.}}};
			\node at (3.2,27/40) {\color{red} \textsf{\textbf{\small{$e_{1jgl}$}}}};	
			\node at (3.2,37/40) {\color{red} \textsf{\textbf{\scriptsize{Share}}}};	
			
			\node [rotate=25] at (3.6,-0.6) {\color{blue} \textsf{\textbf{\tiny{$HF_1(K_s)$}}}};		
			\draw [thick,blue,dashed] (4,0) -- (4,6/40);
			\node at (4,6/40) {\color{blue} \textbf{\Huge{.}}};
			\node at (5.6,6/40) {\color{blue} \textsf{\textbf{\small{$s\_in_{jkl}=HE_1(d_{jkl})$}}}};
			\node at (5.6,16/40) {\color{blue} \textsf{\textbf{\scriptsize{Inner signature}}}}; 
			
			\node [rotate=25] at (4.6,-0.6) {\color{green} \textsf{\textbf{\tiny{$HF_1(ID_4)$}}}};	
			\draw [thick,green,dashed] (5,0.7) -- (5,101/40);
			\node at (5,101/40) {\color{green} \textbf{\Huge{.}}};
			\node at (3.9,101/40) {\color{green} \textsf{\textbf{\scriptsize{$HE_2(pk_{jk},ID_4))$}}}};	
			\node at (3.9,111/40) {\color{green} \textsf{\textbf{\scriptsize{Pseudo share}}}};	
			
			\node [rotate=25] at (5.6,-0.6) {\color{green} \textsf{\textbf{\tiny{$HF_1(ID_5)$}}}};
			\draw [thick,green,dashed] (6,0.7) -- (6,99/40);
			\node at (6,99/40) {\color{green} \textbf{\Huge{.}}};
			\node at (7,110/40) {\color{green} \textsf{\textbf{\scriptsize{$HE_2(pk_{jk},ID_5))$}}}};
			\node at (7,120/40) {\color{green} \textsf{\textbf{\scriptsize{Pseudo share}}}};	
			
			\node [rotate=25] at (7.6,-0.6) {\color{red} \textsf{\textbf{\tiny{$HF_1(ID_3)$}}}};	
			\draw [thick,red,dashed] (8,0) -- (8,40/40);
			\node at (8,40/40) {\color{red} \textbf{\Huge{.}}};
			\node at (8,50/40) {\color{red} \textsf{\textbf{\small{$e_{3jgl}$}}}};	
			\node at (8,60/40) {\color{red} \textsf{\textbf{\scriptsize{Share}}}};	
			
			\node at (-0.25,0.5) {\textsf{\scriptsize{20}}};
			\node at (-0.25,1.0) {\textsf{\scriptsize{40}}};
			\node at (-0.25,1.5) {\textsf{\scriptsize{60}}};
			\node at (-0.25,2.0) {\textsf{\scriptsize{80}}};
			\node at (-0.31,2.5) {\textsf{\scriptsize{100}}};
			\node at (-0.31,3.0) {\textsf{\scriptsize{120}}};
			\node at (-0.31,3.5) {\textsf{\scriptsize{140}}};
			\draw [thick] (0,0.5) -- (-0.05,0.5);
			\draw [thick] (0,1.0) -- (-0.05,1.0);
			\draw [thick] (0,1.5) -- (-0.05,1.5);
			\draw [thick] (0,2.0) -- (-0.05,2.0);
			\draw [thick] (0,2.5) -- (-0.05,2.5);
			\draw [thick] (0,3.0) -- (-0.05,3.0);
			\draw [thick] (0,3.5) -- (-0.05,3.5);	

		\end{tikzpicture}
	} 
	\vspace{-0.5cm}
	\caption{Sample sharing process}
	\label{fig:sample-data-and-shares}
\end{figure}
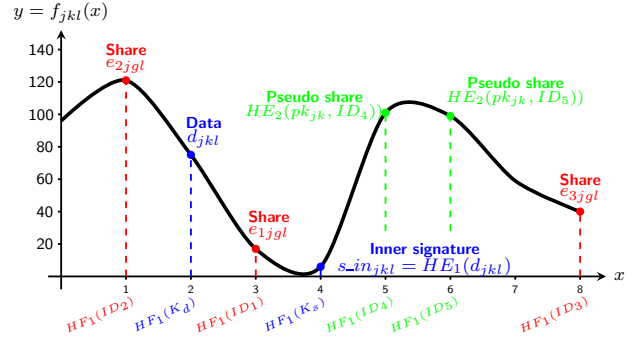

Suppose we want to share a piece of data $d_{jkl}$, e.g., the category ID of product \#124 in Figure \ref{fig:example-DW}(a). We also need to share its inner signature $s\_in_{jkl}$ to enforce data integrity. To generate a polynomial of degree $t$, we need $t-2$ more values that we call pseudo shares. Usually, secret sharing schemes use random polynomials. In contrast, we construct a polynomial by Lagrange interpolation using $d_{jkl}$, $s\_in_{jkl}$ and the $t-2$ pseudo shares. Then, we can share $d_{jkl}$ and $s\_in_{jkl}$ at $n-t+2$ CSPs. Figure \ref{fig:sample-data-and-shares} plots an example where $t=4$. Shares $e_{1jgl}$, $e_{2jgl}$ and $e_{3jgl}$ are created from polynomial $f_{jkl}(x)$ of degree $t-1=3$.

To reconstruct $d_{jkl}$, let us assume we select the following set of $t$ CSPs: $RG=\lbrace CSP_1,CSP_2,CSP_4,CSP_5\rbrace$. Then, if $e_{ijgl}$ is stored at $CSP_i$, it is used for reconstruction. Otherwise, the corresponding pseudo share is used instead. To ease this operation, bitmaps representing where shares are stored are maintained in the index server(s). For example, the bitmap corresponding to product \#124 in Figure \ref{fig:example-DW} is $10101$, with a $1$ value at position $\#i$ representing share storage at $CSP_i$.

The remainder of this section details the sharing and reconstruction processes (Section~\ref{sec:fVSS}), our novel outer signatures (Section~\ref{sec:Outer-signatures}) and the way we share data warehouses (Section~\ref{sec:DWdesign}), achieve loading, backup and recovery processes (Section~\ref{sec:Load-backup-recovery}) and perform OLAP operations (Section~\ref{sec:Data-analysis}).

\subsection{Data Sharing and Reconstruction} 
\label{sec:fVSS}


In fVSS, DB attribute values, except NULL values and primary or foreign keys, are encrypted and shared in relational DBs at CSPs'. Keys help match records in the data reconstruction process and perform join and grouping operations. Any sensitive primary key, such as a social security number, is replaced by an unencrypted sequential integer key. 

Each value $d_{jkl}$ of attribute $A_{jl}$ in record $R_{jk}$ from table $T_j$ is encrypted into $n-t+2$ shares generated from a polynomial $f_{jkl}(x)$ of degree $t-1$ created by Lagrange interpolation from $d_{jkl}$, its inner signature $s\_in_{jkl}$, the identifier numbers $ID_i$ of the CSPs selected to store the shares (this set of CSPs is denoted $SG_{jk}$), and two generated keys $K_d$ and $K_s$ for data and signatures, respectively. The inner signature we use is very similar to that of \cite{DB:Attasena-et-al-2013}. It is created from $d_{jkl}$ by homomorphic encryption \cite{Homomorphism}. $s\_in_{jkl}$ matches with $d_{jkl}$ in the reconstruction process only if CSPs in $RG$ return correct shares. 

\subsubsection{Initialization Phase}
\begin{enumerate}
\item	Set values of $p$ (a big prime number), $n$ and $t$.
\item	Define one-variable hash function $HF_1(a)$ where $a$ is an integer and hash values $HF_1(a)$ must be small integers.
\item	Define one-variable homomorphic function $HE_1(h)$ such that $HE_1(h)$ and $h$ are reals and $HE_1(h_1) \pm HE_1(h_2)=HE_1(h_1 \pm h_2)$.
\item	Define two-variable homomorphic function $HE_2(a,b)$, where $HE_2(a,b)$, $a$ and $b$ are reals and $HE_2(a_1,b)+HE_2(a_2,b)=HE_2(a_1+a_2,b)$.
\item	Set values of CSP identifiers $ID_{i=1..n}$, $K_d$ and $K_s$ such that their values range in $]0, p[$. All $HF_1(ID_i)$ must be unique and different from $HF_1(K_d)$ and $HF_1(K_s)$.
\end{enumerate}

\subsubsection{Data Sharing Process}
Any record $R_{jk}$ is encrypted independently as follows.
\begin{enumerate}
\item Determine the group of CSPs $SG_{jk}$ that will store $R_{jk}$'s $n-t+2$ shares. Let $UG_{jk}$ be the group of CSPs that do not store $R_{jk}$'s shares, i.e., $UG_{jk} = \lbrace CSP_i \rbrace_{i=1..n}- SG_{jk}$.
\item For each attribute $A_{jl}$:
	\begin{enumerate}
	\item Compute $d_{jkl}$'s inner signature: \\$s\_in_{jkl}=HE_1(d_{jkl})$.
	\item Create polynomial $f_{jkl}(x)$ of degree $t-1$ by Lagrange interpolation (Equation~\ref{eq:polynomial equation}):
	\begin{equation}
	f_{jkl}(x)=\sum_{\alpha = 1}^t \prod _{1\leq \beta \leq t,\alpha \neq\beta} \frac{x-x_\beta}{x_\alpha - x_\beta}\times y_\alpha
	\label{eq:polynomial equation}
	\end{equation}
	where $\lbrace (x_1,y_2),\dots ,(x_t,y_t) \rbrace = \\
	\lbrace(HF_1(K_d),d_{jkl}),(HF_1(K_s),s\_in_{jkl})_{CSP_i\in SG_{jk}}\rbrace \\
	\cup \lbrace (HF_1(ID_i),HE_2(pk_{jk},ID_i))_{CSP_i\in UG_{jk}} \rbrace$. \\
	$(HF_1(ID_i),HE_2(pk_{jk},ID_i))$ are pseudo shares.
	\item Compute the set of $d_{jkl}$'s $n-t+2$ shares $\left\{ e_{ijgl}\right\}$. $\forall CSP_i\in SG_{jk}$: $e_{ijgl}=f_{jkl}\left(HF_1(ID_i)\right)$, with $pk_{jk}=pk_{ijg}$.
	\end{enumerate}
\end{enumerate}
  
Following this routine, record $R_{jk}$ is shared into $n-t+2$ records $ER_{ijg}$ at CSPs in $SG_{jk}$. The relationship between $R_{jk}$ and $ER_{ijg}$ is maintained through primary keys $pk_{jk}=pk_{ijg}$. Finally, the bitmap corresponding to $R_{jk}$ is stored in the index server(s) at this time, knowing $SG_{jk}$ and $UG_{jk}$.

Finally, since each data piece is shared independently, it is easy to handle the usual data types featured in DBs. Integers, dates and timestamps can be directly shared by fVSS. Data of other types (i.e., reals, characters, strings and binary strings) are first transformed into integers before being shared \cite{DB:Attasena-et-al-2013}.

\subsubsection{Data Reconstruction Process}
\label{sec:rec}

Any attribute value $d_{jkl}$ is reconstructed as follows. 

\begin{enumerate}
\item Select $t$ CSPs to form reconstruction group $RG$.
\item For each $CSP_i \in RG$, if outer data verification (Section \ref{sec:Outer-signatures}) outputs an error, replace $CSP_i$ by another CSP selected from $\{CSP_i\}_{i=1..n}-RG$.
\item For each $CSP_i\in SG_{jk}\cap RG$, load share $e_{ijgl}$ into $y_i$ where $pk_{jk}=pk_{ijg}$.
\item For each $CSP_i\in UG_{jk}\cap RG$, compute pseudo share $y_i=HE_2(pk_{jk},ID_i)$.
\item Create polynomial $f_{jkl}(x)$ of degree $t-1$ (Equation \ref{eq:polynomial equation}) with $x_i=HF_1(ID_i)$.
\item Compute value $d_{jkl}=f_{jkl}(HF_1(K_d))$.
\item Compute inner signature $s\_in_{jkl}=f_{jkl}(HF_1(K_s))$.
\item Verify $d_{jkl}$'s correctness: if $s\_in_{jkl} \neq HE_1(d_{jkl})$, then restart reconstruction process at step \#1 with a new $RG$.
\end{enumerate}

\subsubsection{Recapitulative Example}
\label{sec:Example}

Let us refer back to Figure~\ref{fig:sample-data-and-shares}, where $n=5$ and $t=4$. The set of CSPs selected for sharing an attribute value $d_{jkl}$ is $SG_{jk}=\lbrace CSP_1,CSP_2,CSP_3\rbrace$. Thus, $UG_{jk}=\lbrace CSP_4,CSP_5\rbrace$, from which pseudo shares $HE2(pk_{jk}, ID_i)_{i \in \{4, 5\}}$, where $pk_{jk}$ is the primary key value of record $R_{jk}$, are computed. Polynomial $f_{jkl}(x)$ is created by Lagrange interpolation from $d_{jkl}$, its inner signature $s\_in_{jkl}$, pseudo shares and keys $K_d$ and $K_s$. Then shares of $d_{jkl}$ are: $e_{ijgl}=f_{jkl}(H1(ID_i))_{i \in \{1, 2, 3\}}$.

Assuming the set of $t$ CSPs selected for reconstruction is $RG=\lbrace CSP_1,CSP_2,CSP_4,CSP_5\rbrace$, $d_{jkl}$ is reconstructed from shares $e_{ijgl~i \in \{1, 2\}}$ (since $CSP_1, CSP_2 \in SG_{jk}$) and pseudo shares $HE2(pk_{jk}, ID_i)_{i \in \{4, 5\}}$ (since $CSP_4, CSP_5 \in UG_{jk}$).



\subsection{Outer Signatures}
\label{sec:Outer-signatures}

Outer data verification helps determine whether data integrity is compromised by CSPs (willingly or not). For this sake, we propose two new types of outer signatures: record and table signatures (whereas \cite{DB:Attasena-et-al-2013} uses an attribute value-level signature). 

\begin{figure}[htb]
	\resizebox{85mm}{!} {
		\begin{tikzpicture}
		
		\node [rectangle,fill=black!70,minimum size=5mm,minimum width=60mm] at (-3,3.2) {\color{white} \textsf{\textbf{\scriptsize{Table $ET_{i,1}$}}}};
		\node [rectangle,fill=black!70,minimum size=5mm,minimum width=60mm] at (-3,2.5) {\color{white} \textsf{\textbf{\scriptsize{Table $ET_{i,2}$}}}};
		\node at (-1.4,1.8) { \textsf{\textbf{\huge{$\vdots$}}}};
		\node [rectangle,fill=black!70,minimum size=5mm,minimum width=60mm] at (-3,0.7) {\color{white} \textsf{\textbf{\scriptsize{Table $ET_{i,m-1}$}}}};
		\node [rectangle,fill=black!70,minimum size=5mm,minimum width=60mm] at (-3,0.0) {\color{white} \textsf{\textbf{\scriptsize{Table $ET_{i,m}$}}}};
		
		\node [rectangle,draw,thick,minimum size=4mm,minimum width=60mm] at (-3,-0.5) (data) {\textbf{\small{$\ $}}};
		\node at (-5.25,-0.5) { \textsf{\textbf{\scriptsize{P.Keys}}}};
		\node at (-3.75,-0.5) { \textsf{\textbf{\scriptsize{$A_{m1}$}}}};
		\node at (-2.25,-0.5) { \textsf{\textbf{\scriptsize{$\cdots$}}}};
		\node at (-0.75,-0.5) { \textsf{\textbf{\scriptsize{$A_{mq_m}$}}}};
		\node [rectangle,draw,thick,minimum size=4mm,minimum width=60mm] at (-3,-0.9) (data) {\textbf{\small{$\ $}}};
		\node at (-5.25,-0.9) { \textsf{\textbf{\scriptsize{$pk_{im1}$}}}};
		\node at (-3.75,-0.9) { \textsf{\textbf{\scriptsize{$e_{im11}$}}}};
		\node at (-2.25,-0.9) { \textsf{\textbf{\scriptsize{$\cdots$}}}};
		\node at (-0.75,-0.9) { \textsf{\textbf{\scriptsize{$e_{im1q_m}$}}}};
		\node [rectangle,draw,thick,minimum size=6mm,minimum width=60mm] at (-3,-1.4) (data) {\textbf{\small{$\ $}}};
		\node at (-5.25,-1.3) { \textsf{\textbf{\scriptsize{$\vdots$}}}};
		\node at (-3.75,-1.3) { \textsf{\textbf{\scriptsize{$\vdots$}}}};
		\node at (-2.25,-1.3) { \textsf{\textbf{\scriptsize{$\ddots$}}}};
		\node at (-0.75,-1.3) { \textsf{\textbf{\scriptsize{$\vdots$}}}};
		\node [rectangle,draw,thick,minimum size=4mm,minimum width=60mm] at (-3,-1.9) (data) {\textbf{\small{$\ $}}};
		\node at (-5.25,-1.9) { \textsf{\textbf{\scriptsize{$pk_{imer_{im}}$}}}};
		\node at (-3.75,-1.9) { \textsf{\textbf{\scriptsize{$e_{imer_{im}1}$}}}};
		\node at (-2.25,-1.9) { \textsf{\textbf{\scriptsize{$\cdots$}}}};
		\node at (-0.75,-1.9) { \textsf{\textbf{\scriptsize{$e_{imer_{im}q_m}$}}}};
		\draw [thick] (-1.5,-0.3) -- (-1.5,-2.1);
		\draw [thick] (-3.0,-0.3) -- (-3.0,-2.1);
		\draw [thick] (-4.5,-0.3) -- (-4.5,-2.1);
		
		\node [circle,fill=black!100,minimum size=1mm] at (2.3,3.2) (nT1) {};
		\draw [thick,dashed] (1.0,3.4) -- (nT1);
		\draw [thick,dashed] (1.0,3.0) -- (nT1);
		\node at (2.5,3.5) {$s\_tout_{i01}$};
		
		\node [circle,fill=black!100,minimum size=1mm] at (2.3,2.5) (nT2) {};
		\draw [thick,dashed] (1.0,2.7) -- (nT2);
		\draw [thick,dashed] (1.0,2.3) -- (nT2);
		
		\node [circle,fill=black!100,minimum size=1mm] at (2.3,1.8) (nT3) {};
		\draw [thick,dashed] (1.0,2.0) -- (nT3);
		\draw [thick,dashed] (1.0,1.6) -- (nT3);
		
		\node [circle,fill=black!100,minimum size=1mm] at (2.3,0.7) (nT4) {};
		\draw [thick,dashed] (1.0,0.9) -- (nT4);
		\draw [thick,dashed] (1.0,0.5) -- (nT4);
		
		\node [circle,fill=black!100,minimum size=1mm] at (0.3,-0.9) (n1) {};
		\node [circle,fill=black!100,minimum size=1mm] at (0.3,-1.3) (n2) {};
		\node [circle,fill=black!100,minimum size=1mm] at (0.3,-1.9) (n3) {};
		\node [circle,fill=black!100,minimum size=1mm] at (1.3,-1.1) (n4) {};
		\node [circle,fill=black!100,minimum size=1mm] at (2.3,-1.4) (nTm) {};
		\draw [thick] (n1) -- (n4);
		\draw [thick,dashed] (n2) -- (n4);
		\draw [thick,dashed] (n3) -- (nTm);
		\draw [thick,dashed] (n4) -- (nTm);
		\node at (0.9,-0.4) {$s\_rout_{im01}$};
		\node at (1.8,-0.8) {$s\_rout_{im11}$};
		\node at (1.1,-2.3) {$s\_rout_{im0r_m}$};
		\node at (3.5,-1.4) {$s\_tout_{i0m}$ or};
		\node at (4,-1.7) {$s\_rout_{imur_m}$ where};
		\node at (3.7,-2.1) {$u=\lceil \log_wr_m \rceil$};
		
		\node [circle,fill=black!50,minimum size=1mm] at (3.3,2.85) (nT5) {};
		\node [circle,fill=black!50,minimum size=1mm] at (3.3,2.15) (nT6) {};
		\node [circle,fill=black!50,minimum size=1mm] at (3.3,-0.35) (nT7) {};
		\draw [thick,gray] (nT1) -- (nT5);
		\draw [thick,gray] (nT2) -- (nT5);
		\draw [thick,gray,dashed] (nT3) -- (nT6);
		\draw [thick,gray] (nT4) -- (nT7);
		\draw [thick,gray] (nTm) -- (nT7);
		\node [circle,fill=black!50,minimum size=1mm] at (4.3,2.5) (nT8) {};
		\node [circle,fill=black!50,minimum size=1mm] at (4.3,0.1) (nT9) {};
		\node [circle,fill=black!50,minimum size=1mm] at (5.3,1.3) (nT10) {};
		\draw [thick,gray] (nT5) -- (nT8);
		\draw [thick,gray,dashed] (nT6) -- (nT8);
		\draw [thick,gray,dashed] (nT7) -- (nT9);
		\draw [thick,gray,dashed] (nT8) -- (nT10);
		\draw [thick,gray,dashed] (nT9) -- (nT10);
		\node at (3.7,3.1) {$s\_tout_{i11}$};
		\node at (4.7,2.8) {$s\_tout_{i21}$};
		\node at (5.8,0.9) {$s\_tout_{iv1}$};
		\node at (5.8,0.6) {where};
		\node at (5.8,0.2) {$v=\lceil \log_wm \rceil$};

		\draw [thick,gray,dashed] (2.3,-1.7) -- (2.3,-2.9);
		\node at (1.2,-2.9) {$\underbrace{\hspace{60pt}}$};
		\node at (1.2,-3.15) {Record signature};
		\node at (1.2,-3.40) {trees};
		\node at (4.5,-2.9) {$\underbrace{\hspace{120pt}}$};
		\node at (4.5,-3.15) {Table signature};
		\node at (4.5,-3.40) {tree};
		\node at (3.4,-3.7) {$\underbrace{\hspace{180pt}}$};
		\node at (3.4,-3.95) {Outer signature tree};
		
		\node [fill=black!5,minimum size=15mm,minimum width=50mm] at (-3,-3.15) (nT5) {};
		\node [circle,fill=black!100,minimum size=1mm] at (-4,-2.9){};
		\node at (-2.6,-2.9) { \textsf{\textbf{\scriptsize{Record signature}}}};
		\node [circle,fill=black!50,minimum size=1mm] at (-4,-3.4){};
		\node at (-2.7,-3.4) { \textsf{\textbf{\scriptsize{Table Signature}}}};

		\end{tikzpicture}
	} 
	\vspace{-0.5cm}
	\caption{Outer signature tree at $CSP_i$}
	\label{fig:my-DB-framework-outer-Signatures}
\end{figure}
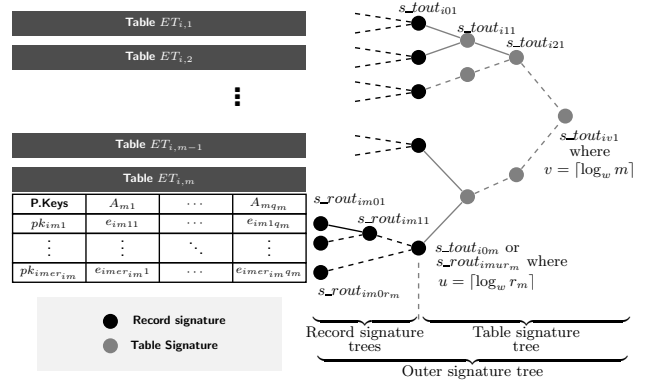

Moreover, we also propose a tree data structure to efficiently exploit outer signatures (Figure \ref{fig:my-DB-framework-outer-Signatures}). Signature trees are stored at CSPs'. The maximum number of child nodes in the signature tree at $CSP_i$ is denoted $w_i$. Each signature tree is constituted of two subtrees: a table signature tree and record signature subtree. Leaf nodes of the record signature tree are record signatures. Higher-level nodes represent the signatures of record clusters, until the root, which is a table signature and thus a leaf of the table signature tree. Similarly, higher level nodes represent the signatures of table groups, until the root, which stores the whole DB's signature.



Signatures are checked before reconstructing data (Section \ref{sec:rec}). Data integrity can also be verified on-demand. 
Outer record signatures are created with the help of user-defined one-way functions $HF_{*i}(h)$. Shared record $ER_{ijg}$ passes integrity check if $s\_rout_{ij0g}=HF_{*i}(ER_{ijg})$.
 
Similarly, shared table $ET_{ij}$ passes integrity check if $s\_tout_{i0j}$ $= HE_{*i}(\sum _{g=1}^{er_{ij}} HF_{*i}(ER_{ijg}))$, where $HE_{*i}(h)$ are homomorphic functions and $er_{ij}$ is the number of records in $ET_{ij}$.  
Thanks to the homomorphism property \cite{Homomorphism}, all tables are correct if a signature stored in a root node equals the sum of the signatures stored in all its child nodes. This allows directly checking a set of records or tables through one node only, and thus speeds up data integrity verification. Moreover, in case of error, we can also discover where the integrity breach is by testing whether each node respects $s\_out_{iuv}=HE_{*i}(\sum _{a=(v-1)\times w_i}^{v\times w_i} s\_out_{i(u-1)a})$, where $s\_out_{iuv}$ may either be a record or a table outer signature, $u > 0$ is a level number and $v$ a node number in level $u$. 

Finally, note that the signature tree's root level is $u_r = \lceil \log _{w_i}m\rceil+\lceil \log _{w_i}\max (er_{ij})_{j=1\cdots m}\rceil -1$, the table signature tree's leaf level is $u_t = u_r -\lceil \log _{w_i}m\rceil$ and the record signature tree's leaf level is $u_t -\lceil \log _{w_i} er_{ij} \rceil +1$. 



\subsubsection{Setup}
For each $CSP_i$, the following parameters must be user-defined.
\begin{enumerate}
\item	Determine $w_i$.
\item	Define one-way function $HF_{*i}(ER_{ijg})$.
\item	Define homomorphic function $HE_{*i}(h)$, where \\$HE_{*i}(h_1) \pm HE_{*i}(h_2)=HE_{*i}(h_1 \pm h_2)$.
\end{enumerate}

\subsubsection{Shared Table Creation}
Whenever a new table $ET_{ij}$ is created at $CSP_i$, the table signature tree is updated from leaf to root as follows.
\begin{enumerate}
\item	Compute $ET_{ij}$'s table signature $s\_tout_{i0j}=HF_{*i}(0)$ and store it in a new right-most leaf node of the table signature tree.
\item	Recursively create new parent nodes $(u,v)$ up to the root of the table signature tree such that\\ $u=1\dots\lceil \log _{w_i}j\rceil$ and $v=\lceil j/(w_i)^u\rceil$.
	\begin{enumerate}
	\item If the right-most node at level $u$ bears the maximum number of children, i.e., $v=(j-1) \mod (w_i)^u$, insert a new right-most parent node $(u,v)$ with value $s\_tout_{i0j}$. 
	\item If there is no node at level $u$ such that $j=(w_i)^{u-1}$ $+1$, insert a new root node $(u,1)$ with value \\$s\_tout_{iu1}=s\_tout_{i(u-1)1}$. 
 	\item Otherwise, stop recursion. 
	\end{enumerate}
\end{enumerate}

\subsubsection{Shared Record Insertion}
Whenever a new record $ER_{ijg}$ is inserted into shared table $ET_{ij}$, the outer signature tree is updated from leaf to root as follows.
\begin{enumerate}
\item	Compute record signature $s\_rout_{ij0g}=HF_{*i}(ER_{ijg})$ and store it in a new right-most leaf node of $ET_{ij}$'s record signature tree.
\item	Recursively insert or update parent nodes $(u,v)$ up to the root of $ET_{ij}$'s record signature tree such that $u=1\dots\lceil \log _{w_i}g\rceil$ and $v=\lceil g/(w_i)^u\rceil$.
	\begin{enumerate}
	\item If the right-most node at level $u$ bears the maximum number of children, i.e., $v=(g-1) \mod (w_i)^u$, insert a new right-most parent node $(u,v)$ with value $s\_rout_{ij0g}$. 
	\item If there is no node in level $u$ such that $g=(w_i)^{u-1}$ $+1$, insert a new root node $(u,1)$ with value \\$s\_rout_{iu1}=HE_{*i}(s\_rout_{ij(u-1)1}+s\_rout_{ij0g})$. \\Compute signature change before and after insertion: $\Delta=HE_{*i}(s\_rout_{ij(u-1)1}-s\_rout_{ij0g})$.
 	\item If root node has been reached, compute signature change before and after insertion: \\$\Delta=HE_{*i}(s\_rout_{ijuv}-s\_rout_{ij0g})$. Then, update the root node's value to $HE_{*i}(s\_rout_{ijuv}+s\_rout_{ij0g})$.
 	\item Otherwise, update the parent node's signature $s\_rout_{iuv}$ with $HE_{*i}(s\_rout_{ijuv}+s\_rout_{ij0g})$.
	\end{enumerate}
\item	Starting from the root of $ET_{ij}$'s record signature tree, which is also leaf $s\_tout_{i(0)j}$ of the corresponding table signature tree, recursively update parent nodes up to the root of the table signature tree such that $u=1\dots\lceil \log _{w_i}m\rceil$ and $v=\lceil j/(w_i)^u\rceil$. Then, update table signature $s\_tout_{iuv}$ to $HE_{*i}(s\_tout_{iuv}+\Delta)$.
\end{enumerate}

\begin{figure}[htb]
	\resizebox{85mm}{!} {
		\begin{tikzpicture}
	\node [circle,fill=black!100,minimum size=1mm] at (-4.8,0.5)(xD01){};
	\node at (-4.8,0.25) {\textsf{\tiny{1}}};
	\node [circle,fill=black!100,minimum size=1mm] at (-4.4,0.5)(xD02){};
	\node at (-4.4,0.25) {\textsf{\tiny{2}}};
	\node [circle,fill=black!100,minimum size=1mm] at (-4.0,0.5)(xD03){};
	\node at (-4.0,0.25) {\textsf{\tiny{3}}};
	
	\node [circle,fill=black!100,minimum size=1mm] at (-3.4,0.5)(xD04){};
	\node at (-3.4,0.25) {\textsf{\tiny{4}}};
	\node [circle,fill=black!100,minimum size=1mm] at (-3.0,0.5)(xD05){};
	\node at (-3.0,0.25) {\textsf{\tiny{5}}};
	\node [circle,fill=black!100,minimum size=1mm] at (-2.6,0.5)(xD06){};
	\node at (-2.6,0.25) {\textsf{\tiny{6}}};
	
	\node [circle,fill=black!100,minimum size=1mm] at (-2.0,0.5)(xD07){};
	\node at (-2.0,0.25) {\textsf{\tiny{7}}};
	\node [circle,fill=black!100,minimum size=1mm] at (-1.6,0.5)(xD08){};
	\node at (-1.6,0.25) {\textsf{\tiny{8}}};
	\node [circle,fill=black!100,minimum size=1mm] at (-1.2,0.5)(xD09){};
	\node at (-1.2,0.25) {\textsf{\tiny{9}}};
		
	\node [circle,fill=black!100,minimum size=1mm] at (-4.4,1)(xD11){};
	\draw[thick] (xD01) -- (xD11);
	\draw[thick] (xD02) -- (xD11);
	\draw[thick] (xD03) -- (xD11);
	\node [circle,fill=black!100,minimum size=1mm] at (-3.0,1)(xD12){};
	\draw[thick] (xD04) -- (xD12);
	\draw[thick] (xD05) -- (xD12);
	\draw[thick] (xD06) -- (xD12);
	\node [circle,fill=black!100,minimum size=1mm] at (-1.6,1)(xD13){};
	\draw[thick] (xD07) -- (xD13);
	\draw[thick] (xD08) -- (xD13);
	\draw[thick] (xD09) -- (xD13);
	
	\node [circle,fill=black!60,minimum size=1mm] at (-5,1.5)(xT01){};
	\node at (-4.75,1.5) {\textsf{\tiny{1}}};
	\node [circle,fill=black!60,minimum size=1mm] at (-4,1.5)(xT02){};
	\node at (-3.75,1.5) {\textsf{\tiny{2}}};
	\node [circle,fill=black!60,minimum size=1mm] at (-3,1.5)(xT03){};
	\node at (-2.75,1.5) {\textsf{\tiny{3}}};
	\draw[thick] (xD11) -- (xT03);
	\draw[thick] (xD12) -- (xT03);
	\draw[thick] (xD13) -- (xT03);
	
	\node [circle,fill=black!60,minimum size=1mm] at (-0.8,1.5)(xT04){};
	\node at (-0.55,1.5) {\textsf{\tiny{4}}};
	
	\node [circle,fill=black!60,minimum size=1mm] at (-4,2.0)(xT11){};
	\draw[thick] (xT01) -- (xT11);
	\draw[thick] (xT02) -- (xT11);
	\draw[thick] (xT03) -- (xT11);
	
	\node [circle,fill=black!60,minimum size=1mm] at (-0.8,2.0)(xT12){};
	\draw[thick] (xT04) -- (xT12);
	
	\node [circle,fill=black!60,minimum size=1mm] at (-2.4,2.5)(xT21){};
	\draw[thick] (xT11) -- (xT21);
	\draw[thick] (xT12) -- (xT21);

	\node at (0,1.5) {\textsf{\huge{$\Rightarrow$}}};
	
	\node [circle,fill=black!100,minimum size=1mm] at (0.0,0)(D01){};
	\node at (0.0,-0.25) {\textsf{\tiny{1}}};
	\node [circle,fill=black!100,minimum size=1mm] at (0.4,0)(D02){};
	\node at (0.4,-0.25) {\textsf{\tiny{2}}};
	\node [circle,fill=black!100,minimum size=1mm] at (0.8,0)(D03){};
	\node at (0.8,-0.25) {\textsf{\tiny{3}}};
	
	\node [circle,fill=black!100,minimum size=1mm] at (1.4,0)(D04){};
	\node at (1.4,-0.25) {\textsf{\tiny{4}}};
	\node [circle,fill=black!100,minimum size=1mm] at (1.8,0)(D05){};
	\node at (1.8,-0.25) {\textsf{\tiny{5}}};
	\node [circle,fill=black!100,minimum size=1mm] at (2.2,0)(D06){};
	\node at (2.2,-0.25) {\textsf{\tiny{6}}};
	
	\node [circle,fill=black!100,minimum size=1mm] at (2.8,0)(D07){};
	\node at (2.8,-0.25) {\textsf{\tiny{7}}};
	\node [circle,fill=black!100,minimum size=1mm] at (3.2,0)(D08){};
	\node at (3.2,-0.25) {\textsf{\tiny{8}}};
	\node [circle,fill=black!100,minimum size=1mm] at (3.6,0)(D09){};
	\node at (3.6,-0.25) {\textsf{\tiny{9}}};
	
	\node [circle,draw,thick,minimum size=1mm] at (4.2,0)(D010){};
	\node at (4.2,-0.25) {\textsf{\tiny{10}}};
	\node at (5.2,0) {\textsf{\tiny{$s\_dout_{i(0)(10)}$}}};
		
	\node [circle,fill=black!100,minimum size=1mm] at (0.4,0.5)(D11){};
	\draw[thick] (D01) -- (D11);
	\draw[thick] (D02) -- (D11);
	\draw[thick] (D03) -- (D11);
	
	\node [circle,fill=black!100,minimum size=1mm] at (1.8,0.5)(D12){};
	\draw[thick] (D04) -- (D12);
	\draw[thick] (D05) -- (D12);
	\draw[thick] (D06) -- (D12);
	
	\node [circle,fill=black!100,minimum size=1mm] at (3.2,0.5)(D13){};
	\draw[thick] (D07) -- (D13);
	\draw[thick] (D08) -- (D13);
	\draw[thick] (D09) -- (D13);
	
	\node [circle,draw,thick,minimum size=1mm] at (4.2,0.5)(D14){};
	\draw[ultra thick] (D010) -- (D14);
	\node at (5.2,0.5) {\textsf{\tiny{$s\_dout_{i(1)(4)}$}}};
		
	\node [circle,fill=black!100,minimum size=1mm] at (1.8,1.0)(D21){};
	\draw[thick] (D11) -- (D21);
	\draw[thick] (D12) -- (D21);
	\draw[thick] (D13) -- (D21);
		
	\node [circle,draw,thick,minimum size=1mm] at (4.2,1.0)(D22){};
	\draw[ultra thick] (D14) -- (D22);
	\node at (5.2,1.0) {\textsf{\tiny{$s\_dout_{i(2)(2)}$}}};
	
	\node [circle,draw,thick,minimum size=1mm] at (3.0,1.5)(D31){};
	\draw[thick] (D21) -- (D31);
	\draw[ultra thick] (D22) -- (D31);
	\node at (4.0,1.4) {\textsf{\tiny{$s\_dout_{i(3)(1)}$}}};
	\node at (4.0,1.6) {\textsf{\tiny{$s\_tout_{i(0)(3)}$}}};
	
	\node [circle,fill=black!60,minimum size=1mm] at (1,1.5)(T01){};
	\node at (0.75,1.5) {\textsf{\tiny{1}}};
	\node [circle,fill=black!60,minimum size=1mm] at (2,1.5)(T02){};
	\node at (1.75,1.5) {\textsf{\tiny{2}}};
	
	\node [circle,fill=black!60,minimum size=1mm] at (5.2,1.5)(T04){};
	\node at (5.45,1.5) {\textsf{\tiny{4}}};
	
	\node [circle,draw,thick,minimum size=1mm] at (2,2.0)(T11){};
	\node at (3,2.0) {\textsf{\tiny{$s\_tout_{i(2)(1)}$}}};
	\draw[thick] (T01) -- (T11);
	\draw[thick] (T02) -- (T11);
	\draw[ultra thick] (D31) -- (T11);
	
	\node [circle,fill=black!60,minimum size=1mm] at (5.2,2.0)(T12){};
	\draw[thick] (T04) -- (T12);
	
	\node [circle,draw,thick,minimum size=1mm] at (3.6,2.5)(T21){};
	\node at (4.6,2.5) {\textsf{\tiny{$s\_tout_{i(3)(1)}$}}};
	\draw[ultra thick] (T11) -- (T21);
	\draw[thick] (T12) -- (T21);

		\end{tikzpicture}
	} 
	\vspace{-0.5cm}
	\caption{Sample outer signature tree update}
	\label{fig:example-update-tree}
\end{figure}
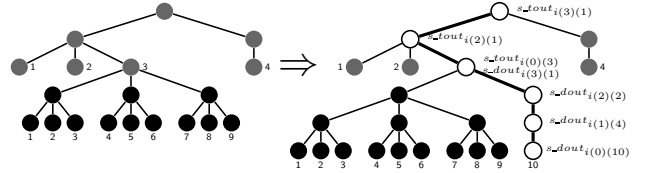

An example of outer signature tree update is provided in Figure \ref{fig:example-update-tree}. Record $ER_{i(3)(10)}$ is a new record; all white nodes are updated or inserted. 

\subsubsection{Shared Record Update}
Whenever a record $ER_{ijg}$ is updated in shared table $ET_{ij}$, the outer signature tree is updated from leaf to root as follows.
\begin{enumerate}
\item	Compute signature change before and after insertion: $\Delta=HE_{*i}(s\_rout_{ij0g}-HF_{*i}(ER_{ijg})$.
\item	Recursively update nodes $(u,v)$ up to the root of $ET_{ij}$'s record signature tree such that $u=0\dots\lceil \log _{w_i}er_{ij}\rceil$ and $v=\lceil g/(w_i)^u\rceil$. Update $s\_rout_{ijuv}$ to \\$HE_{*i}(s\_rout_{ijuv}+\Delta)$.
\item	Starting from the root of $ET_{ij}$'s record signature tree, recursively update parent nodes up to the root of the table signature tree such that $u=1\dots\lceil \log _{w_i}m\rceil$ and $v=\lceil j/(w_i)^u\rceil$. Update table signature $s\_tout_{iuv}$ to $HE_{*i}(s\_tout_{iuv}+\Delta)$.
\end{enumerate}

\subsection{Sharing Data Warehouses}
\label{sec:DWdesign}

Since each table of a shared DW is stored in a relational database at a given CSP's and each attribute value in each record is encrypted independently, our approach straightforwardly helps implement any DW logical model, i.e., star, snowflake and constellation schemas. A shared DW bears the same schema as the original DW's. However, all encrypted attributes are transformed into reals by the sharing process (Section~\ref{sec:fVSS}).


Finally, to improve query performance, computation and data transfer costs when performing Relational OLAP (ROLAP) operations, we use indices and cloud cubes, which are described below.


\subsubsection{Indices}
We exploit three types of indices, which are all created at data-sharing time. In addition to so-called Type~I indices used in the reconstruction process (Section \ref{sec:fVSS}), Type~II and III indices are specifically aimed at enhancing query performance and thus computation cost.

Type~II indices are customarily used by the second family of secret sharing approaches (Section~\ref{sec:Related works} \cite{DB:Hadavi-et-al-2013,DB:Hadavi-et-al-2012,DB:Wang-et-al-2011}). They allow computing exact match, range  and aggregation (e.g., MAX, MIN, MEDIAN and COUNT) queries, without reconstructing data, with the help of B++ trees. Type~II indices are independently stored in the index server(s).
 
Type~III indices help compute aggregate functions such as variance and standard deviation, as well as multiplications and divisions between two attributes, without reconstructing data. 
Type~III indices are stored as extra attributes in shared tables. Let us illustrate why they are needed through examples. To compute the variance and standard deviation over an attribute $X$, $X^2$ values are needed. Then, either we reconstruct (i.e., decrypt) all values of $X$, or we share $X^2$ values in a new attribute, i.e., a Type~III index, and computation can operate directly on shares. Similarly, computing SUM($X \times Y$) or SUM($X \div Y$) requires sharing $X \times Y$ and $X \div Y$ as Type~III indices, respectively, because homomorphism properties only work for summation and subtraction.

\subsubsection{Cloud Cubes}
As in \cite{DB:Attasena-et-al-2013}, our approach supports the storage of data cubes that optimize response time and bandwidth when performing ROLAP operations. In addition, in this work, cubes are directly created in the cloud and refreshed through shares and indices only.

Since cloud cubes are built from shares, they are physically stored into tables that must be shared at all $n$ CSPs, because pseudo shares are not available. In addition to customary dimension references and aggregate measures, they can include additional attributes that actually are embedded Type~III indices. For example, suppose we need to compute the average of measure $M$ from a cube. Then SUM($M$) and COUNT($M$) must be stored in the cube too, to allow computations on shares without reconstruction.


Figure~\ref{fig:Cube1} features a cloud cube named Cube-I that sums total prices and numbers of sales by time period and by product. As is customary, NULL values are used to encode superaggregates. All aggregate measures can be queried directly from Cube-I without reconstruction. 

\begin{figure}[htb]
	\resizebox{80mm}{!} {
	\begin{tabular}{|c|c|c|c|c|c|c|}
	\multicolumn{5}{c}{{\bfseries Foreign Keys}} \\
	\multicolumn{5}{c}{$\overbrace{\hspace{250pt}}$} & 
	\multicolumn{2}{c}{\bfseries (Shares)}\\
	
	\multicolumn{3}{c}{{\bfseries Time Attributes}} & 
	\multicolumn{2}{c}{{\bfseries Product Attributes}} & 
	\multicolumn{2}{c}{{\bfseries Aggregation Attributes}}\\
	\multicolumn{3}{c}{$\overbrace{\hspace{140pt}}$} & 
	\multicolumn{2}{c}{$\overbrace{\hspace{110pt}}$} & 
	\multicolumn{2}{c}{$\overbrace{\hspace{120pt}}$}\\
    \hline
    {\bfseries YearID} & {\bfseries MonthID} & {\bfseries DateID} & 
    {\bfseries CategoryID} & {\bfseries ProdNo} &
    {\bfseries TotalPrice} & {\bfseries Number} \\ \hline
    
    \textit{NULL} & \textit{NULL} & \textit{NULL} & \textit{NULL} & \textit{NULL} & 83231 & 58244\\ \hline
    \textit{NULL} & \textit{NULL} & \textit{NULL} & 1 & \textit{NULL} & 26701 & 18254\\ \hline
    \textit{NULL} & \textit{NULL} & \textit{NULL} & 1 & 1 & 8958 & 7113\\ \hline
    \textit{NULL} & \textit{NULL} & \textit{NULL} & 1 & $\vdots$ & $\vdots$ & $\vdots$\\ \hline
    \textit{NULL} & \textit{NULL} & \textit{NULL} & 1 & 2 & 4348 & 1844\\ \hline
    \textit{NULL} & \textit{NULL} & \textit{NULL} & $\vdots$ & $\vdots$ & $\vdots$ & $\vdots$\\ \hline
    1 & \textit{NULL} & \textit{NULL} & \textit{NULL} & \textit{NULL} & 44574 & 54542\\ \hline
    1 & \textit{NULL} & \textit{NULL} & 1 & \textit{NULL} & 21158 & 8954\\ \hline
    1 & \textit{NULL} & \textit{NULL} & 1 & 1 & 9754 & 4544\\ \hline
    1 & \textit{NULL} & \textit{NULL} & 1 & $\vdots$ & $\vdots$ & $\vdots$\\ \hline
    1 & \textit{NULL} & \textit{NULL} & 2 & 1 & 18444 & 5747\\ \hline
    1 & \textit{NULL} & \textit{NULL} & $\vdots$ & $\vdots$ & $\vdots$ & $\vdots$\\ \hline
    1 & 1 & \textit{NULL} & \textit{NULL} & \textit{NULL} & 8312 & 5812\\ \hline 
    1 & 1 & \textit{NULL} & 1 & \textit{NULL} & 2312 & 1822\\ \hline 
    1 & 1 & \textit{NULL} & 1 & 1 & 988 & 586 \\ \hline 
    1 & 1 & \textit{NULL} & 1 & $\vdots$  & $\vdots$  & $\vdots$ \\ \hline 
    1 & 1 & \textit{NULL} & 2 & 1 & 756 & 458 \\ \hline 
    1 & 1 & \textit{NULL} & $\vdots$  & $\vdots$  & $\vdots$  & $\vdots$ \\ \hline 
    1 & 2 & \textit{NULL} & \textit{NULL} & \textit{NULL} & 9758 & 6254\\ \hline 
    1 & $\vdots$ & \textit{NULL} & $\vdots$ & $\vdots$ & $\vdots$ & $\vdots$\\ \hline 
    1 & 1 & 1 & \textit{NULL} & \textit{NULL} & 2578 & 1587\\ \hline  
    1 & 1 & 1 & 1 & \textit{NULL} & 548 & 425\\ \hline 
    1 & 1 & 1 & 1 & 1 & 56 & 24\\ \hline 
    1 & 1 & 1 & 1 & $\vdots$ & $\vdots$ & $\vdots$ \\ \hline 
    1 & 1 & 1 & 1 & 2 & 95 & 67\\ \hline 
    1 & 1 & 1 & 1 & $\vdots$ & $\vdots$ & $\vdots$ \\ \hline 
    1 & 1 & 1 & 2 & \textit{NULL} & 689 & 357 \\ \hline 
    1 & 1 & $\vdots$ & $\vdots$ & $\vdots$ & $\vdots$ & $\vdots$ \\ \hline 
    $\vdots$ & $\vdots$ & $\vdots$ & $\vdots$ & $\vdots$ & $\vdots$ & $\vdots$ \\ \hline
    \end{tabular}
	} 
	\caption{Sample cloud cube Cube-I}
	\label{fig:Cube1}
\end{figure}

\subsection{Backup, Recovery and Load processes}
\label{sec:Load-backup-recovery}

In our approach, as in all secret sharing approaches, backups are unnecessary because each shared table $ET_{ij}$ is actually a backup share of all other shared table $ET_{ia}$, where $a \in{1,\dots,j-1,j+1,\dots,n}$. In case shared tables or shared records are detected as erroneous by outer signature verification (Section~\ref{sec:Outer-signatures}), their can be recovered from $t$ other shared tables. 

Loading new data into an existing shared DW does not require decrypting previous data first, because each attribute value in each record is encrypted independently. For instance, in Figure~\ref{fig:load-DW}, data from Figure~\ref{fig:example-DW} are already shared and the last record (\#127) is new. 
After each new shared record is loaded, the outer signature tree must be updated (Section~\ref{sec:Outer-signatures}), and indices and possible cloud cubes refreshed. 

\begin{figure}[htb]
\begin{center}
\subfigure[Original data]{ 
   \resizebox{4cm}{!} {
    \begin{tabular}{|c|c|c|c|c|}
    \hline
    {\bfseries ProNo} & {\bfseries ProName} & {\bfseries ProDescr} & {\bfseries CategoryID} & {\bfseries UnitPrice} \\ \hline   
    124 & Shirt & Red & 1 & 75 \\ \hline
    125 & Shoe & \textit{NULL} & 2 & 80 \\ \hline
    126 & Ring & \textit{NULL} & 1 & 80 \\ \hline
    127 & Hat & \textit{NULL} & 3 & 5 \\ \hline
    \end{tabular}
    } 
 }\subfigure[Type I indices]{
   \resizebox{4cm}{!} {
    \begin{tabular}{c|c|c|c}
    \cline{2-3}
    {$\hspace{75pt}$} & {\bfseries ProNo} & {\bfseries Share location} & {$\hspace{75pt}$}\\ \cline{2-3}  
    & 124 & 10101 & \\ \cline{2-3}
    & 125 & 01110 & \\ \cline{2-3}
    & 126 & 11010 & \\ \cline{2-3}
    & 127 & 00111 & \\ \cline{2-3}
    \end{tabular}
    }
 }
 \subfigure[Shares at $CSP_{1}$]{
   \resizebox{4cm}{!} {
    \begin{tabular}{|c|c|c|c|c|}
    \hline
    {\bfseries ProNo} & {\bfseries ProName} & {\bfseries ProDescr} & {\bfseries CategoryID} & {\bfseries UnitPrice} \\ \hline  
    124 & \{6,5,3,11,7\} & \{10,5,8\} & 1 & 6 \\ \hline
    126 & \{10,3,6,12\} & \textit{NULL} & 2 & 45 \\ \hline
    \end{tabular}
    }
 }\subfigure[Shares at $CSP_{2}$]{
   \resizebox{4cm}{!} {
    \begin{tabular}{|c|c|c|c|c|}
    \hline
    {\bfseries ProNo} & {\bfseries ProName} & {\bfseries ProDescr} & {\bfseries CategoryID} & {\bfseries UnitPrice} \\ \hline  
    125 & \{6,5,4,5\} & \textit{NULL} & 2 & 5 \\ \hline
    126 & \{2,6,11,10\} & \textit{NULL} & 6 & 8 \\ \hline
    \end{tabular}
    }
 }
 \subfigure[Shares at $CSP_{3}$]{
   \resizebox{4cm}{!} {
    \begin{tabular}{|c|c|c|c|c|}
    \hline
    {\bfseries ProNo} & {\bfseries ProName} & {\bfseries ProDescr} & {\bfseries CategoryID} & {\bfseries UnitPrice} \\ \hline  
    124 & \{6,6,5,7,9\} & \{12,8,1\} & 4 & 7 \\ \hline
    125 & \{6,5,8,3\} & \textit{NULL} & 9 & 11 \\ \hline
    127 & \{7,2,9\} & \textit{NULL} & 5 & 2 \\ \hline
    \end{tabular}
    }
 }\subfigure[Shares at $CSP_{4}$]{
   \resizebox{4cm}{!} {
    \begin{tabular}{|c|c|c|c|c|}
    \hline
    {\bfseries ProNo} & {\bfseries ProName} & {\bfseries ProDescr} & {\bfseries CategoryID} & {\bfseries UnitPrice} \\ \hline  
    125 & \{9,15,13,8\} & \textit{NULL} & 12 & 7 \\ \hline
    126 & \{2,7,6,9\} & \textit{NULL} & 12 & 1 \\ \hline
    127 & \{13,8,3\} & \textit{NULL} & 4 & 9 \\ \hline
    \end{tabular}
    }
 }
 \subfigure[Shares at $CSP_{5}$]{
   \resizebox{4cm}{!} {
    \begin{tabular}{|c|c|c|c|c|}
    \hline
    {\bfseries ProNo} & {\bfseries ProName} & {\bfseries ProDescr} & {\bfseries CategoryID} & {\bfseries UnitPrice} \\ \hline
    
    124 & \{5,9,11,1,5\} & \{10,6,7\} & 8 & 13 \\ \hline
    127 & \{7,17,4\} & \textit{NULL} & 2 & 9 \\ \hline
    \end{tabular}
    }
 }
\end{center}
\vspace{-0.5cm}
\caption{Example of sharing new data} \label{fig:load-DW}
\end{figure}





Cloud cube refreshment requires further attention. When updating cubes, aggregates can be updated from shares and indices. 
There are three cases.

For aggregations by, e.g., MAX and MIN operators, the primary key of aggregates is discovered from a Type~II index. Then, at each CSP's, the share corresponding to the primary key is updated in the cloud cube.  In case no corresponding share is found at that CSP's, the shared cube is updated from a pseudo share with the help of a Type~I index.  

For COUNTs, aggregates can be easily found from a Type~II index and reconstructed. Then, the aggregates can be updated and shared back into the cloud cube \cite{SSS:Shamir-1979}.

For SUMs, thanks to the homomorphism property, aggregates can be updated from cloud cubes by summing shares and pseudo shares.
For example, to update SUM($X$) at $CSP_i$, the new aggregate is the sum of shares and pseudo shares (Equation~\ref{eq:sumX}, where SUM$_{CSP_i}(X)$ is trivially computed at $CSP_i$ and $SUM_{index}(PK_i)$ is computed with the help of a Type~I index). 

	\begin{equation}
		\begin{array}{rr}
			SUM(X)=SUM_{CSP_i}(X)+HE_2(SUM_{index}(PK_i),ID_i) 
		\end{array}
		\label{eq:sumX}
	\end{equation}

Finally, more complex aggregations require combining the above cases. For example, SUM($X \pm Y$) is updated from shares and pseudo shares by Equation~\ref{eq:summationXY}.
	\begin{equation}
		\begin{array}{rl}
			SUM(X \pm Y)= & SUM_{CSP_i}(X+Y) \\
			& +2\times HE_2(SUM_{index}(PK_i),ID_i)
		\end{array}
		\label{eq:summationXY}
	\end{equation}


\subsection{Querying a Shared Data Warehouse}
\label{sec:Data-analysis}

Simple SELECT/FROM queries directly apply onto shares, as well as summing positive integer attributes. All join operators, when operating on unencrypted keys, also apply directly. When expressing conditions in a WHERE or HAVING clause, Type II indices must be used \cite{DB:Hadavi-et-al-2013,DB:Hadavi-et-al-2012,DB:Wang-et-al-2011}. Almost all comparison operators (=, $\neq$, EXISTS, IN, $>$, $\geq$, <, $\leq$, BETWEEN...) can be evaluated against such B++ index trees. 

Similarly, aggregation functions such as MAX, MIN and COUNT can directly apply on shares with the help of Type~II indices \cite{DB:Hadavi-et-al-2013,DB:Hadavi-et-al-2012,DB:Wang-et-al-2011}. 
In contrast, a SUM must combine relevant aggregates of shares and pseudo shares (as in Equation~\ref{eq:sumX}) with an external program before reconstruction. 
	
Other aggregation functions must be computed by an external program after reconstructing relevant aggregates from shares and pseudo shares. Average, variance and standard deviation are computed as in Equations~\ref{eq:avgX}, \ref{eq:varX} and \ref{eq:sdX}, respectively, where $X^2$ is a Type III index.

	\begin{equation}
		\begin{array}{rr}
			AVG(X)=DC(SUM(X))/DC(COUNT(X)) 
		\end{array}
		\label{eq:avgX}
	\end{equation}
	\begin{equation}
		\begin{array}{rr}
			VAR(X)=\frac{DC(SUM(X^2))}{DC(COUNT(X))}+\frac{DC(SUM(X))^2}{DC(COUNT(X))^2} 
		\end{array}
		\label{eq:varX}
	\end{equation}
	\begin{equation}
		\begin{array}{rr}
			STDDEV(X)=\sqrt{\frac{DC(SUM(X^2))}{DC(COUNT(X))}+\frac{DC(SUM(X))^2}{DC(COUNT(X))^2}} 
		\end{array}
		\label{eq:sdX}
	\end{equation}

When aggregating calculated fields, multiplication and division can be performed directly from Type III indices. However, summation and subtraction between two attributes must combine relevant aggregates of shares and pseudo shares with an external program before reconstruction. Aggregates at $CSP_i$ compute as in Equation~\ref{eq:summationXY}.

Finally, grouping queries using the GROUP BY or GROUP BY CUBE clauses can directly apply on shares if they target unencrypted key attributes. Again, grouping by other attribute(s) requires the use of a Type II index.

Consequently, executing some queries may require either transforming or splitting the query, depending on its clauses and operators, following the above guidelines. Figure~\ref{fig:Ex-complex-query} shows a sample query and the way it runs at the user's, at one index server and at $CSP_1$ (query processing at other index servers and CSPs is similar). For clarity, the user, index server and CSP are denoted U, IS and CSP, respectively in Figure~\ref{fig:Ex-complex-query}.

\begin{figure}[htb]
\centering
    \label{tab:Ex-complex-query}
    
	\subfigure[Original query]{ 
		\resizebox{8.35cm}{!} {
   		\begin{tikzpicture}
   		\node [rectangle,draw,thick] at (0,0){
   		\begin{tabular}{l}
    		SELECT SUM(S.price+S.tax) AS sumprice, P.prodName \\
    		FROM Sale AS S JOIN Product AS P ON S.ProdNo=P.ProdNo \\
    		JOIN Date AS D ON S.DateKey=D.DateKey \\
    		WHERE D.Date BETWEEN '2014-01-01' AND '2014-01-15' \\
    		GROUP BY P.prodName 
    	\end{tabular} };
    	\end{tikzpicture}
    	}
   	}
   	\subfigure[Execution steps]{ 
		\resizebox{8.35cm}{!} {
   		\begin{tikzpicture}
   		\node [rectangle,draw,thick] at (0,0){
    	\begin{tabular}{p{8.35cm}}
    		1. (IS) Match DateKey in Type II index with condition \\
    		~~~ D.Date BETWEEN '2014-01-01' AND '2014-01-15'. \\
    		~~~ Let $DK$ be the resulting set. \\ [4pt]    	
    		2. (U) Query-I is created to run at each CSP's: \\
    		~~~ SELECT P.prodNo, P.prodName, \\
    		~~~ SUM(S.price+S.tax) AS sumprice \\
    		~~~ FROM Sale AS S JOIN Product AS P \\
    		~~~ ON S.ProdNo=P.ProdNo \\
    		~~~ WHERE S.Datekey IN ($DK$) \\
    		~~~ GROUP BY P.ProdNo \\ [4pt]
    		3. (CSP) Execute Query-I. W.r.t. Figure~\ref{fig:load-DW}, result is:  \\
    		~~~ $\{~\lbrace 124,\lbrace 6,5,3,11,7 \rbrace,789\rbrace$, \\ 
    		~~~~~~ $\lbrace 126,\lbrace 10,3,6,12 \rbrace,945\rbrace~\}$.   \\ [4pt]  		
    		4. (U) Query-II is created to run on Type I index: \\
    		~~~ SELECT prodNo, SUM(OrderNo) AS sumPK, \\
    		~~~ FROM Sale \\
    		~~~ WHERE Datekey IN ($DK$) \\
    		~~~ GROUP BY ProdNo\\ [4pt]
    		5. (IS) Execute Query-II.
    	 Let us denote the results \\
    	 ~~~ $sumPK_{pi}$, where $sumPK_{pi}$ is 
    		 sumPK's value for \\
    	~~~ product 
    	 $p$ and its $CSP_i$ pseudo share.     \\ [4pt]  				
    		6. (U) Reconstruct each record (prodName, sumprice) \\
    		~~~ in the query result. For example, to reconstruct \\
    		~~~ record prodNo=124: \\
    		~~~ $CSP_1$ share of prodName is $\lbrace 6,5,3,11,7 \rbrace$ and \\
    		~~~ $CSP_1$ aggregate of sumprice is \\
    		~~~ $789+2\times HE_2(sumPK_{(124)(1)},ID_1)$ (Equation~\ref{eq:summationXY}).
    	\end{tabular} };
    	\end{tikzpicture}
    	}
    }
    \vspace{-0.5cm}    
    \caption{Sample query rewriting}
	\label{fig:Ex-complex-query}
\end{figure}

Finally, our approach directly supports all basic OLAP operations by directly querying cloud cubes and reconstructing the global result. For example, the total price and the number of products per year can be queried from Cube-I by query: \\
{\small
"SELECT YearID, YearName, TotalPrice, Number \\
FROM Cube-I, year \\
WHERE Cube-I.YearID = year.YearID AND MonthID IS NULL AND DateID IS NULL AND CategoryID IS NULL AND Prod-No IS NULL"}.\\
To drill down to total price and number of products per month in 2014, the previous query becomes: \\
{\small
"SELECT YearID, YearName, \emph{Month}, TotalPrice, Number \\
FROM Cube-I, year, month \\
WHERE Cube-I.YearID = year.YearID \emph{AND Cube-I.MonthID = Month.MonthID AND Cube-I.YearID = 2014} AND DateID IS NULL AND CategoryID IS NULL AND Prod-No IS NULL"}.



\section{Comparative Study}
\label{sec:Comparison-approaches}

In this section, we compare fVSS to the related approaches presented in Section~\ref{sec:Related works}, with respect to security and cost in the pay-as-you-go paradigm, global cost being customarily divided into storage, computing and data transfer costs. Table~\ref{tab:Comparison-approaches} synthesizes the features and complexities of all approaches, which we discuss below.

\begin{table*}[tb]
\centering
    \caption{Comparison of database sharing approaches}     
    \label{tab:Comparison-approaches}
    \begin{small}
    \resizebox{177mm}{!} {
    \begin{tabular}{|l|c|c|c|c|c|c|c|c|c|}
    \hline
    \multicolumn{1} {|c|} {\bfseries Features and costs} & 
    {\bfseries\cite{DB:Agrawal-et-al-2009}} & {\bfseries\cite{DB:Attasena-et-al-2013}} & 
    {\bfseries\cite{DB:Emekci-et-al-2005}} & {\bfseries\cite{DB:Emekci-et-al-2006}} & 
    {\bfseries\cite{DB:Hadavi-et-al-2013}} & {\bfseries\cite{DB:Hadavi-Jalili-2010,DB:Hadavi-et-al-2012}} & 
    {\bfseries\cite{DB:Thompson-et-al-2009}} & {\bfseries\cite{DB:Wang-et-al-2011}} &  
    {\bfseries fVSS} \\
    \hline
    
    Data privacy &
    Yes & Yes & Yes & Yes & Yes & Yes & Yes & Yes & Yes \\ 
    \hline
    
    Data availability & Yes & Yes & Yes & Yes & Yes & Yes & Yes & Yes & Yes \\ 
    \hline
    
    Ability in case CSPs& & & & & & & & & \\ 
    fail, to& & & & & & & & & \\ 
    - Query shares & 
    Yes & Yes & Yes & Yes & Yes & Yes & Yes & Yes & Yes \\     
    - Update shares &
    No & No & No & No & No & No & No & No & Yes  \\ 
    \hline
    
    Data integrity & & & & & & & & & \\ 
    - Inner code verifying &
    No & Yes & No & No & No & No & Yes & Yes & Yes \\ 
    - Outer code verifying &
    No & Yes & No & No & No & No & No & No & Yes \\ 
    \hline
    
    Target &
    DBs & DWs & DWs & DBs & DBs & DBs & DBs & DBs & DWs \\ 
    \hline
    
    Data sources &
    Single & Single & Multi & Multi & Single & Single & Single & Single & Single \\ 
    \hline
    
    Data types &
    Positive & Integers, & Positive & Integers & Integers & Positive &  Positive & Positive & Integers, \\  
     & integers & Reals, & integers & & & integers & integers & integers & Reals, \\ 
     & & Characters, & & & & & & & Characters, \\ 
     & & Strings, & & & & & & & Strings, \\
     & & Dates, & & & & & & & Dates, \\
     & & Booleans & & & & & & & Booleans \\
    \hline
    
    Shared data access & & & & & & & & & \\ 
    - Updates &
    No & Yes & No & Yes & Yes & Yes & Yes & Yes & Yes \\  
    - Exact match queries &
    No & Yes & Yes & Yes & Yes & Yes & No & Yes & Yes \\ 
    - Range queries &
    No & No & Yes & Yes & Yes & Yes & No & Yes & Yes \\ 
    - Aggregation queries &
    Yes & Yes & Yes & Yes & Yes & Yes & Yes & No & Yes \\ 
    $\ \ $ on one attribute & & & & & & & & & \\ 
    - Aggregation queries &
    No & No & No & No & No & No & No & No & Yes \\ 
    $\ \ $ on two attributes & & & & & & & & & \\ 
    - Grouping queries &
    No & Yes & No & No & No & No & No & No & Yes \\ 
    \hline

    Complexity & & & & & & & & & \\ 
    - Data storage w.r.t. &
    $\geq n$ & $\geq n/\left(t-1\right)$ & $\geq 2n$ & $\geq 2n$ & $\geq n$ & $\geq n$ & $\geq 2n$ & $\geq n/t$ & $\geq n-t+2$ \\
    $\ \ $ original data volume & & + signatures & & & & +1 (B++ tree) & +1 (hash tree) & +$n/t$ (B++ tree) & +1 (B++ tree) \\ 
    & & & & & & & & + signatures & + signatures \\ 
    - Sharing time &
    $O\left(\sigma nt\right)$ & $O\left(\sigma nt\right)$ & $O\left(\sigma nt\right)$ & $O\left(\sigma nt\right)$ & $O\left(\sigma nt\right)$ & $O\left(\sigma nt\right)$ & $O\left(\sigma nt\right)$ & $O\left(\max\left(\sigma\log \sigma ,\sigma n\right)\right)$ & $O\left(\sigma t\left(n-t\right)\right)$ \\ 
    - Reconstruction time &
    $O\left(\gamma t^{2}\right)$ & $O\left(\gamma t^{2}\right)$ & $O\left(\gamma t^{2}\right)$ & $O\left(\gamma t^{2}\right)$ & $O\left(\gamma t^{2}\right)$ & $O\left(\gamma t^{2}\right)$ & $O\left(\gamma t^{2}\right)$ & $O\left(\gamma t\right)$ & $O\left(\gamma t^{2}\right)$ \\ 
    \hline
    \end{tabular}
    }
    \end{small} 
\end{table*}

\subsection{Data Security Features}

By data security, we mean data privacy, availability and integrity. By design, all secret sharing-based approaches enforce privacy by guaranteeing shares cannot be decrypted by a single CSP or an intruder who would hack a CSP. Actually, a coalition or the compromise  of at least $t$ CSPs is necessary to break the secret. Privacy is further improved in fVSS, because data is not shared at all CSPs', but only $n-t+2$. Thence, fVSS imposes a new constraint: no CSP \emph{group} can hold enough shares to reconstruct the original data if $n < 2 \times t - 2$. Indeed, $n < 2 \times t - 2 \Leftrightarrow n-t+2<t$, i.e., the number of shares is lower than the number of shares necessary for reconstruction.

With respect to availability, all secret sharing-based approa-ches, still by design, allow reconstructing the secret, i.e., query shares, when $n-t$ CSPs fail. However, to the best of our knowledge, only fVSS allows updating shares in case of CSP failure(s), simply by not selecting the failing CSP(s) for sharing new data. This is again possible because data is shared at $n-t+2$ CSPs instead of $n$.

Finally, to enforce integrity, only \cite{DB:Attasena-et-al-2013} and fVSS verify both the correctness of shares and the honesty of CSPs by outer and inner code verification, respectively. Only two other approaches use inner code verification alone. The novelty of fVSS is that outer signatures are computed at the record and table granularity, instead of the attribute value's, which allows faster verification, e.g., when checking one table signature instead all signatures of one or several attributes in said table.

\subsection{Storage Cost}

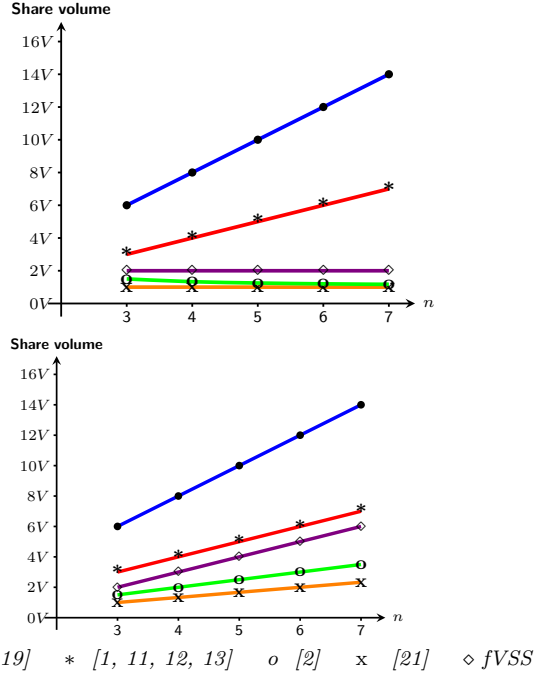
\begin{figure}[htb]
\begin{center}
	\resizebox{60mm}{!} {
		\begin{tikzpicture}
			\draw[-stealth,thick] (-0.2,0) -- (5.4,0) node[right] {\textsf{\textbf{\scriptsize{$n$}}}};
			\draw[-stealth,thick] (0,-0.2) -- (0,4.3) node[above] {\textsf{\textbf{\scriptsize{Share volume}}}};
			
			\draw[color=black,ultra thick,blue] plot [smooth] coordinates {(1,6/4) (2,8/4) (3,10/4) (4,12/4) (5,14/4)};
			\node at (1,6/4) {\textbf{\Huge{.}}};
			\node at (2,8/4) {\textbf{\Huge{.}}};
			\node at (3,10/4) {\textbf{\Huge{.}}};
			\node at (4,12/4) {\textbf{\Huge{.}}};
			\node at (5,14/4) {\textbf{\Huge{.}}};
					
			\draw[color=black,ultra thick,red] plot [smooth] coordinates { (1,3/4) (2,4/4) (3,5/4) (4,6/4) (5,7/4)};
			\node at (1,3/4) {\textbf{{*}}};
			\node at (2,4/4) {\textbf{{*}}};
			\node at (3,5/4) {\textbf{{*}}};
			\node at (4,6/4) {\textbf{{*}}};
			\node at (5,7/4) {\textbf{{*}}};
			
			\draw[color=black,ultra thick,orange] plot [smooth] coordinates { (1,0.25) (2,0.25) (3,0.25) (4,0.25) (5,0.25)};
			\node at (1,0.25) {\textbf{{x}}};
			\node at (2,0.25) {\textbf{{x}}};
			\node at (3,0.25) {\textbf{{x}}};
			\node at (4,0.25) {\textbf{{x}}};
			\node at (5,0.25) {\textbf{{x}}};
			
			\draw[color=black,ultra thick,green] plot [smooth] coordinates { (1,1.5/4) (2,1.33/4) (3,1.25/4) (4,1.2/4) (5,1.17/4)};
			\node at (1,1.5/4) {\textbf{{o}}};
			\node at (2,1.33/4) {\textbf{{o}}};
			\node at (3,1.25/4) {\textbf{{o}}};
			\node at (4,1.2/4) {\textbf{{o}}};
			\node at (5,1.17/4) {\textbf{{o}}};
			
			\draw[color=black,ultra thick,violet] plot [smooth] coordinates { (1,0.5) (2,0.5) (3,0.5) (4,0.5) (5,0.5)};
			\node at (1,0.5) {\textbf{{$\diamond$}}};
			\node at (2,0.5) {\textbf{{$\diamond$}}};
			\node at (3,0.5) {\textbf{{$\diamond$}}};
			\node at (4,0.5) {\textbf{{$\diamond$}}};
			\node at (5,0.5) {\textbf{{$\diamond$}}};

			\node at (1,-0.2) {\textsf{\scriptsize{3}}};
			\node at (2,-0.2) {\textsf{\scriptsize{4}}};
			\node at (3,-0.2) {\textsf{\scriptsize{5}}};
			\node at (4,-0.2) {\textsf{\scriptsize{6}}};
			\node at (5,-0.2) {\textsf{\scriptsize{7}}};
			\draw [thick] (1,0) -- (1,-0.05);
			\draw [thick] (2,0) -- (2,-0.05);
			\draw [thick] (3,0) -- (3,-0.05);
			\draw [thick] (4,0) -- (4,-0.05);
			\draw [thick] (5,0) -- (5,-0.05);
			
			
			\node at (-0.3,0.0) {\textsf{\scriptsize{$0V$}}};
			\node at (-0.3,0.5) {\textsf{\scriptsize{$2V$}}};
			\node at (-0.3,1.0) {\textsf{\scriptsize{$4V$}}};
			\node at (-0.3,1.5) {\textsf{\scriptsize{$6V$}}};
			\node at (-0.3,2.0) {\textsf{\scriptsize{$8V$}}};
			\node at (-0.35,2.5) {\textsf{\scriptsize{$10V$}}};
			\node at (-0.35,3.0) {\textsf{\scriptsize{$12V$}}};
			\node at (-0.35,3.5) {\textsf{\scriptsize{$14V$}}};
			\node at (-0.35,4.0) {\textsf{\scriptsize{$16V$}}};
			\draw [thick] (0,0.5) -- (-0.05,0.5);
			\draw [thick] (0,1.0) -- (-0.05,1.0);
			\draw [thick] (0,1.5) -- (-0.05,1.5);
			\draw [thick] (0,2.0) -- (-0.05,2.0);
			\draw [thick] (0,2.5) -- (-0.05,2.5);
			\draw [thick] (0,3.0) -- (-0.05,3.0);
			\draw [thick] (0,3.5) -- (-0.05,3.5);
			\draw [thick] (0,4.0) -- (-0.05,4.0);	

		\end{tikzpicture}
	}	
	\resizebox{60mm}{!} {
		\begin{tikzpicture}
			\draw[-stealth,thick] (-0.2,0) -- (5.4,0) node[right] {\textsf{\textbf{\scriptsize{$n\hspace{15pt}$}}}};
			\draw[-stealth,thick] (0,-0.2) -- (0,4.3) node[above] {\textsf{\textbf{\scriptsize{Share volume}}}};
			
			\draw[color=black,ultra thick,blue] plot [smooth] coordinates { (1,6/4) (2,8/4) (3,10/4) (4,12/4) (5,14/4)};
			\node at (1,6/4) {\textbf{\Huge{.}}};
			\node at (2,8/4) {\textbf{\Huge{.}}};
			\node at (3,10/4) {\textbf{\Huge{.}}};
			\node at (4,12/4) {\textbf{\Huge{.}}};
			\node at (5,14/4) {\textbf{\Huge{.}}};
					
			\draw[color=black,ultra thick,red] plot [smooth] coordinates { (1,3/4) (2,4/4) (3,5/4) (4,6/4) (5,7/4)};
			\node at (1,3/4) {\textbf{{*}}};
			\node at (2,4/4) {\textbf{{*}}};
			\node at (3,5/4) {\textbf{{*}}};
			\node at (4,6/4) {\textbf{{*}}};
			\node at (5,7/4) {\textbf{{*}}};
			
			\draw[color=black,ultra thick,orange] plot [smooth] coordinates { (1,1/4) (2,1.33/4) (3,1.67/4) (4,2/4) (5,2.33/4)};
			\node at (1,1/4) {\textbf{{x}}};
			\node at (2,1.33/4) {\textbf{{x}}};
			\node at (3,1.67/4) {\textbf{{x}}};
			\node at (4,2/4) {\textbf{{x}}};
			\node at (5,2.33/4) {\textbf{{x}}};
			
			\draw[color=black,ultra thick,green] plot [smooth] coordinates { (1,1.5/4) (2,2/4) (3,2.5/4) (4,3/4) (5,3.5/4)};
			\node at (1,1.5/4) {\textbf{{o}}};
			\node at (2,2/4) {\textbf{{o}}};
			\node at (3,2.5/4) {\textbf{{o}}};
			\node at (4,3/4) {\textbf{{o}}};
			\node at (5,3.5/4) {\textbf{{o}}};
			
			\draw[color=black,ultra thick,violet] plot [smooth] coordinates { (1,2/4) (2,3/4) (3,4/4) (4,5/4) (5,6/4)};
			\node at (1,2/4) {\textbf{{$\diamond$}}};
			\node at (2,3/4) {\textbf{{$\diamond$}}};
			\node at (3,4/4) {\textbf{{$\diamond$}}};
			\node at (4,5/4) {\textbf{{$\diamond$}}};
			\node at (5,6/4) {\textbf{{$\diamond$}}};

			\node at (1,-0.2) {\textsf{\scriptsize{3}}};
			\node at (2,-0.2) {\textsf{\scriptsize{4}}};
			\node at (3,-0.2) {\textsf{\scriptsize{5}}};
			\node at (4,-0.2) {\textsf{\scriptsize{6}}};
			\node at (5,-0.2) {\textsf{\scriptsize{7}}};
			\draw [thick] (1,0) -- (1,-0.05);
			\draw [thick] (2,0) -- (2,-0.05);
			\draw [thick] (3,0) -- (3,-0.05);
			\draw [thick] (4,0) -- (4,-0.05);
			\draw [thick] (5,0) -- (5,-0.05);
			
			
			\node at (-0.3,0.0) {\textsf{\scriptsize{$0V$}}};
			\node at (-0.3,0.5) {\textsf{\scriptsize{$2V$}}};
			\node at (-0.3,1.0) {\textsf{\scriptsize{$4V$}}};
			\node at (-0.3,1.5) {\textsf{\scriptsize{$6V$}}};
			\node at (-0.3,2.0) {\textsf{\scriptsize{$8V$}}};
			\node at (-0.35,2.5) {\textsf{\scriptsize{$10V$}}};
			\node at (-0.35,3.0) {\textsf{\scriptsize{$12V$}}};
			\node at (-0.35,3.5) {\textsf{\scriptsize{$14V$}}};
			\node at (-0.35,4.0) {\textsf{\scriptsize{$16V$}}};
			\draw [thick] (0,0.5) -- (-0.05,0.5);
			\draw [thick] (0,1.0) -- (-0.05,1.0);
			\draw [thick] (0,1.5) -- (-0.05,1.5);
			\draw [thick] (0,2.0) -- (-0.05,2.0);
			\draw [thick] (0,2.5) -- (-0.05,2.5);
			\draw [thick] (0,3.0) -- (-0.05,3.0);
			\draw [thick] (0,3.5) -- (-0.05,3.5);
			\draw [thick] (0,4.0) -- (-0.05,4.0);	

		\end{tikzpicture}
	} 
	 
\small\emph{$\bullet$~~\cite{DB:Emekci-et-al-2005,DB:Emekci-et-al-2006,DB:Thompson-et-al-2009} $\hspace{10pt}*$~~\cite{DB:Agrawal-et-al-2009,DB:Hadavi-et-al-2013,DB:Hadavi-Jalili-2010,DB:Hadavi-et-al-2012} $\hspace{10pt}$o~~\cite{DB:Attasena-et-al-2013}} \small{$\hspace{10pt}$x} \small\emph{~~\cite{DB:Wang-et-al-2011} $\hspace{10pt}\diamond$ fVSS}
\end{center} 	
	\vspace{-0.5cm}
	\caption{Storage complexity comparison}
	\label{fig:storage-cost}
\end{figure}

Storage cost directly depends on shared data volume, which in turn depends on parameters $n$ and $t$. Figure~\ref{fig:storage-cost} plots the volume of shared data for all studied approaches, expressed as a multiple of original data volume $V$, with respect to $n$, with $t=n$ in the upper graph and $t=3$ in the lower graph.  Figure~\ref{fig:storage-cost} shows that fVSS help control shared data volume better than most existing approach, and is close to the best approaches \cite{DB:Attasena-et-al-2013,DB:Wang-et-al-2011} in this respect.


However, storage cost does not only depend on the global volume of shares. Since fVSS allows selecting the data volume shared at each CSP, it can be differentiated to benefit from different pricing policies. Let us illustrate this through an example. Let $n=5$, $t=4$ and $V=100$~GB. CSP pricing policies for this range of data volume are depicted in Table~\ref{tab:CSPs-cost}. Prices are real prices from CSPs such as Amazon web services, Windows azure and Google compute engine.
Finally, let us assume that each individual share is not bigger than the original data it encrypts, e.g., a shared integer is not bigger than the original unencrypted integer. We also disregard index and signature volume, which depends on user-defined parameters in all approaches, for the sake of simplicity.

\begin{table}[htb]
\centering
    \caption{CSP pricing policies}  
    \label{tab:CSPs-cost}
    \resizebox{85mm}{!} {
    \begin{tabular}{|l|c|c|c|c|c|}
    \cline{2-6}
    \multicolumn{1}{c|}{ } & {\bfseries $CSP_1$} & {\bfseries $CSP_2$} & {\bfseries $CSP_3$} & {\bfseries $CSP_4$} & {\bfseries $CSP_5$} \\ \hline
    Storage (\$/GB/month)& 0.030 & 0.040 & 0.053 & 0.120 & 0.325 \\ \hline \hline 
    sVM CPU time (\$/h)& 0.013 & 0.059 & 0.058 & 0.060 & 0.070 \\ \hline  
    mVM CPU time (\$/h) & 0.026 & 0.079 & 0.115 & 0.120 & 0.140 \\ \hline  
    lVM CPU time (\$/h)& 0.053 & 0.120 & 0.230 & 0.240 & 0.280 \\ \hline  
    \end{tabular}
    }
\end{table}

Table~\ref{tab:Comparison-storage-cost} features a storage cost comparison of all studied approaches. The first line relates to unencrypted data stored at one CSP, for reference. Global share volume is computed with respect to data storage complexities (Table~\ref{tab:Comparison-approaches}). Storage cost is the sum of data volumes stored at each $CSP_i$ times $CSP_i$'s storage price. We included two strategies for fVSS. In fVSS-I, data are equitably shared among CSPs. In fVSS-II, data are preferentially shared at $CSP_1$, $CSP_2$ and $CSP_3$, which are the cheapest. Results clearly show that fVSS-II achieves a much lower cost than fVSS-I. Moreover, even though global share volume with fVSS-II is significantly larger that with the most efficient previous approaches \cite{DB:Attasena-et-al-2013,DB:Wang-et-al-2011}, final cost is comparable, and even a little lower. However, \cite{DB:Wang-et-al-2011} is not applicable in our context since it does not allow aggregation queries. It is thus discarded in the following.


\begin{table}[hbt]
\centering
    \caption{Storage cost comparison}  
    \label{tab:Comparison-storage-cost}
    \resizebox{85mm}{!} {
    \begin{tabular}{|c|r|r|r|}
    \hline
    \multirow{2}{*}{\bfseries Approach} & \multicolumn{2}{|c|}{\bfseries Share volume (GB)} & \multirow{2}{*}{\bfseries Storage cost (\$)}  \\ \cline{2-3}   
     & {\bfseries Global} & {\bfseries per CSP} &  \\ \hline
     Unencrypted data & 100 & 100 & 3 to 32.5 \\ \hline
    
    \cite{DB:Emekci-et-al-2005,DB:Emekci-et-al-2006,DB:Thompson-et-al-2009} & 1,000 & 200 & 113.60 \\ \hline
    \cite{DB:Agrawal-et-al-2009,DB:Hadavi-et-al-2013,DB:Hadavi-Jalili-2010,DB:Hadavi-et-al-2012} & 500 & 100 & 56.80 \\ \hline
    \cite{DB:Attasena-et-al-2013} & 167 & 34 & 19.31 \\ \hline
    \cite{DB:Wang-et-al-2011} & 125 & 25 & 14.77 \\ \hline
    fVSS-I & 300 & 60 & 34.08  \\ \hline
    \multirow{2}{*}{fVSS-II} & \multirow{2}{*}{300} & 99.8 + 99.8 + 99.8 & \multirow{2}{*}{\textbf{12.39}} \\ 
    & & + 0.4 + 0.2 & \\ \hline
    \end{tabular}
    }
\end{table}

\subsection{Computing Cost}

\subsubsection{Sharing Cost}
\label{sec:SharingCost}

The sharing process time complexity (Table~\ref{tab:Comparison-approaches}) depends on $n$, $t$ and $\sigma$, which is the number of shared data pieces, i.e., individual attribute values. Since $\sigma$ is normally much bigger than $n$ and $t$, $\sigma \log\sigma>\sigma nt>\sigma t(n-t)>\sigma n$. Moreover, the number of CSPs where records are shared is $n-t+2$ in fVSS and $n$ in all other approaches. If $t \geq 2$, which is quite probable, $n - t + 2 \leq n$. Thus, we expect fVSS to be faster than previous approaches when sharing data.


Let us illustrate this through an example. Let $n=5$, $t=4$ and $\sigma = 10^{15}$. CSP pricing policies are depicted in Table~\ref{tab:CSPs-cost}, where sVM, mVM and lVM stand for small, medium and large virtual machine, respectively. Let us assume that the computing powers of sVMs, mVMs and lVMs are $1\times 10^{10}$, $2\times 10^{10}$ and $4\times 10^{10}$ records per second, respectively. Virtual machine size is assigned to each CSP with respect to the number of records to share at that CSP.

Table~\ref{tab:Comparison-computation-cost-sharing} features a sharing cost comparison of all studied approaches but \cite{DB:Wang-et-al-2011}. Sharing time is the number of processed records divided by the virtual machine's power. CPU cost is the sum of sharing times at each $CSP_i$ times $CSP_i$'s computing price. Results show that both fVSS-I and fVSS-II have a cheapest sharing process than all existing approaches, even though fVSS-I bears the longer sharing time. They also outline again the advantage of unbalancing the volume of shares at CSPs, which helps decrease cost by a factor 2.3 with respect to state-of-the-art approaches in this example.

\begin{table}[hbt]
\centering
    \caption{Sharing cost comparison}  
    \label{tab:Comparison-computation-cost-sharing}
    \resizebox{85mm}{!} {
    \begin{tabular}{|c|r|c|r|r|}
    \hline
    \multirow{2}{*}{\bfseries Approach} & {\bfseries \#records at} & \multirow{2}{*}{\bfseries VM type} & {\bfseries Sharing time} & {\bfseries CPU cost} \\ 
    & {\bfseries each CSP} & & {\bfseries (h:mm)} & {\bfseries (\$)} \\ \hline
    Unencrypted data at 1 CSP & $10^{15}$ & lVM & 6:57 & 0.36 to 1.94 \\ \hline
    \cite{DB:Agrawal-et-al-2009,DB:Attasena-et-al-2013,DB:Emekci-et-al-2005,DB:Emekci-et-al-2006,DB:Hadavi-et-al-2013,DB:Hadavi-Jalili-2010,DB:Hadavi-et-al-2012,DB:Thompson-et-al-2009} & $10^{15}$ & lVM & 6:57 & 6.40  \\ \hline
    fVSS-I & $6\times 10^{14}$ & mVM & 8:20 & 4.40 \\ \hline
    \multirow{5}{*}{fVSS-II} & $9.98\times 10^{14}$ & lVM & 6:56 & \multirow{5}{*}{\textbf{2.80}} \\ 
    & $9.98\times 10^{14}$ & lVM & 6:56 &  \\ 
    & $9.98\times 10^{14}$ & lVM & 6:56 $\Longleftarrow$ 6:56 &  \\ 
    & $4\times 10^{12}$ & sVM & 0:07 &  \\ 
    & $2\times 10^{12}$ & sVM & 0:04 &  \\ \hline
    \end{tabular}
    }
\end{table}

\subsubsection{Data Access Cost}
\label{sec:ReconstructionCost}

Query response time, which is critical in an OLAP context, directly depends on the reconstruction process time complexity, which in turn depends on $t$ and the number of records in the query response, $\gamma$ (Table~\ref{tab:Comparison-approaches}). All secret sharing approaches bear the same reconstruction complexity but \cite{DB:Wang-et-al-2011}. However, this approach cannot compute aggregations on shares, implying all records involved in aggregations must be reconstructed at the user's, which is more costly than computing the aggregation on shares and only reconstruct the result. Moreover, we expect fVSS to be more efficient than previous approaches because we can directly perform all query types on shared DWs and cubes, in parallel, whereas other approaches cannot and must reconstruct bigger datasets before processing them at the user's.



Let us illustrate this through an example. Let $n=5$, $t=4$. Let us assume we run a query matched by 10\% of records, i.e., $\gamma = 10^{14}$, and $RG=\{CSP_1, CSP_2, CSP_4, CSP_5\}$. 
CSP pricing policies and virtual machine power are the same as in Section~\ref{sec:SharingCost}.

Table~\ref{tab:Comparison-computation-cost-querying} features a data access cost comparison of all studied approaches but \cite{DB:Wang-et-al-2011}. Response time is the number of processed records divided by the virtual machine's computing power. CPU cost is the sum of response times at each $CSP_i$ times $CSP_i$'s computing price. Results show again that, even though response time is comparable for all approaches, fVSS allows much lower costs, especially when unbalancing the volume of shares at CSPs in fVSS-II, which helps decrease cost by a factor 4 with respect to state-of-the-art approaches in this example.


\begin{table}[hbt]
\centering
    \caption{Data access cost comparison}  
    \label{tab:Comparison-computation-cost-querying}
    \resizebox{85mm}{!} {
    \begin{tabular}{|c|r|c|r|r|}
    \hline
    \multirow{2}{*}{\bfseries Approach} & {\bfseries \#records at} & \multirow{2}{*}{\bfseries VM type} & {\bfseries Response time} & {\bfseries CPU cost} \\ 
    & {\bfseries each CSP} & & {\bfseries (h:mm)} & {\bfseries (\$)} \\ \hline
    Unencrypted data at 1 CSP & $10^{14}$ & lVM & 0:42 & 0.04 to 0.20 \\ \hline
    \cite{DB:Agrawal-et-al-2009,DB:Attasena-et-al-2013,DB:Emekci-et-al-2005,DB:Emekci-et-al-2006,DB:Hadavi-et-al-2013,DB:Hadavi-Jalili-2010,DB:Hadavi-et-al-2012,DB:Thompson-et-al-2009} & $10^{14}$ & lVM & 0:42 & 0.48  \\ \hline
    fVSS-I & $6\times 10^{13}$ & mVM & 0:50 & 0.30 \\ \hline
    \multirow{5}{*}{fVSS-II} & $9.98\times 10^{13}$ & lVM & 0:42 & \multirow{5}{*}{\textbf{0.12}} \\ 
    & $9.98\times 10^{13}$ & lVM & 0:42 &  \\ 
    & 0~~~ & --- & 0:42 $\Longleftarrow$ 0:00 &  \\ 
    & $4\times 10^{11}$ & sVM & 0:01 &  \\ 
    & $2\times 10^{11}$ & sVM & 0:01 &  \\ \hline
    \end{tabular}
    }
\end{table}

\subsection{Data Transfer Cost}

Data transfer cost directly relates to the size of query results when accessing the shared DW. Since all approaches allow different operations and vary in share volume, it is difficult to compare data transfer cost by proof. However, to reduce data transfer cost, fVSS allows several aggregation operators running on shares. Moreover, by creating shared data cubes, we allow straight computations on shares, and thus only target results are transferred to the user, i.e., with no additional data reconstruction, and thus no stored data transfer.

\section{Conclusion and Perspectives}
\label{sec:conclusion}

In this paper, we propose a new approach for securing cloud DWs, which simultaneously supports data privacy, availability, integrity and OLAP. Our approach builds upon fVSS, which is to the best of our knowledge the first flexible secret sharing that allows users adjusting share volume with respect to CSP pricing polices. Our experiments show that unbalancing share volume at CSPs allows significantly minimizing storage and computing costs in the pay-as-you-go paradigm. Privacy and availability are achieved by design with secret sharing, but fVSS achieves a higher security level and allows DW refreshing even when some CSPs fail. Finally, data integrity is reinforced with both inner and outer signature that help detect errors in query results and shares, respectively.


Future research shall run along three lines. 
First, we plan to further 
assess the cost of our solution in the cloud pay-as-you-go paradigm.  
We especially plan to balance the cost of our solution 
against the cost of risking data loss or theft. 
Moreover, since CSP pricing and servicing policies are likely to evolve quickly, we aim at designing a method for adding and removing CSPs to/from the CSP pool, with the lowest possible update costs and while preserving data integrity.
Second, we aim at designing a tool that semi-automatically helps users adjust the volume of shares at each CSP's, with respect to cost, but also quality of service.
Finally, we also work on share storage management, 
to optimize query performance and reduce both response time and computing cost.

\bibliographystyle{abbrv}
\bibliography{dolap07-attasena}  
 
\balancecolumns

\end{document}